\documentclass[12pt]{article}

\usepackage{lmodern}
\usepackage[utf8]{inputenc}
\usepackage[T1]{fontenc}

\usepackage{amsmath}
\usepackage{graphicx}
\usepackage{color}

\usepackage{amsfonts}
\usepackage{amssymb}
\usepackage{latexsym}
\usepackage{slashed}
\usepackage[linktocpage,unicode]{hyperref}
\usepackage{cite}
\usepackage{mathrsfs}

\newcommand{\tach}{\mathcal{T}}
\newcommand{\itach}{\mathscr{T}}

\newcommand{\eq}[1]{(\ref{#1})}
\newcommand{\ket}[1]{\vert #1 \rangle}

\def\IC{\mathbb{C}}

\def\IZ{\IZ}
\def\IR{\mathbb{R}}
\def\IP{\mathbb{P}}

\def\s{{\text{sgn}}}

\def\re{\mbox{Re }}

\def\tr{\mbox{Tr}}

\def\be{\begin{equation}}
\def\ee{\end{equation}}
\def\bea{\begin{eqnarray}}
\def\eea{\end{eqnarray}}
\def\bes{\begin{subequations}}
\def\ees{\end{subequations}}

\newcommand{\bmat}{\left(\begin{array}}
\newcommand{\emat}{\end{array}\right)}

\def\yzero{\smash{\hbox{$y\kern-4pt\raise1pt\hbox{${}^\circ$}$}}}

\def\beq{\begin{equation}}
\def\eeq{\end{equation}}
\def\beqa{\begin{eqnarray}}
\def\eeqa{\end{eqnarray}}

\def\-{\hphantom{-}}
\def\ov{\overline}
\def\s2{\frac{1}{\sqrt2}}

\def\tr{{\rm tr \,}}

\def\diag{{\rm diag \,}}
\def\IF{\relax{\rm I\kern-.18em F}}
\def\II{\relax{\rm I\kern-.18em I}}

\def\Dsl{\,\raise.15ex\hbox{/}\mkern-13.5mu D} 

\def\IC{{\bf C}}
\def\IS{{\bf S}}
\def\IR{{\bf R}}
\def\IZ{{\bf Z}}
\def\IX{{\bf X}}

\def\IP{\bf P}

\def\re{{\rm Re}\,}

\def\C{{\bf C}}

\newcommand{\cO}{\mathcal{O}}
\newcommand{\bC}{\IC}
\newcommand{\bP}{\IP}
\newcommand{\bR}{\IR}
\newcommand{\bZ}{\IZ}
\newcommand{\sx}{\mathsf{x}}
\newcommand{\sy}{\mathsf{y}}
\newcommand{\sV}{\mathsf{V}}
\newcommand{\sX}{\mathsf{X}}

\catcode`\@=11   
\newdimen\@rotdimen
\newbox\@rotbox  

\def\@vspec#1{\special{ps:#1}}
\def\@rotstart#1{\@vspec{gsave currentpoint currentpoint translate
   #1 neg exch neg exch translate}}
\def\@rotfinish{\@vspec{currentpoint grestore moveto}}
\def\@rotr#1{\@rotdimen=\ht#1\advance\@rotdimen by\dp#1%
   \hbox to\@rotdimen{\hskip\ht#1\vbox to\wd#1{\@rotstart{90 rotate}%
   \box#1\vss}\hss}\@rotfinish}
\def\@rotl#1{\@rotdimen=\ht#1\advance\@rotdimen by\dp#1%
   \hbox to\@rotdimen{\vbox to\wd#1{\vskip\wd#1\@rotstart{270 rotate}%
   \box#1\vss}\hss}\@rotfinish}%
\def\@rotu#1{\@rotdimen=\ht#1\advance\@rotdimen by\dp#1%
   \hbox to\wd#1{\hskip\wd#1\vbox to\@rotdimen{\vskip\@rotdimen
   \@rotstart{-1 dup scale}\box#1\vss}\hss}\@rotfinish}%
\def\@rotf#1{\hbox to\wd#1{\hskip\wd#1\@rotstart{-1 1 scale}%
   \box#1\hss}\@rotfinish}%
\def\rotate{\@ifnextchar[{\@rotate}{\@rotate[l]}}
\def\@rotate[#1]#2{\setbox\@rotbox=\hbox{#2}\@nameuse{@rot#1}\@rotbox}

\catcode`\@=12

\topmargin
-1.5cm
\textwidth
15.5cm
\textheight
23.5cm
\oddsidemargin
0.7cm
\evensidemargin
1.2cm

\begin{document}

\makeatletter
\@addtoreset{equation}{section}
\makeatother
\renewcommand{\theequation}{\thesection.\arabic{equation}}
\pagestyle{empty}
\rightline{MPP-2015-105}
\rightline{IFT-UAM/CSIC-15-053}
\rightline{FTUAM-15-14}
\vspace{0.1cm}
\begin{center}
\LARGE{\bf Closed tachyon solitons\\ in type II string theory\\[12mm]}
\large{Iñaki García-Etxebarria$^1$, Miguel Montero$^{2,3}$, Angel M. Uranga$^2$\\[3mm]}
\footnotesize{$^1$ Max Planck Institute for Physics, 80805 Munich, Germany\\[2mm]
$^2$  Instituto de Física Teórica IFT-UAM/CSIC,\\[-0.3em] 
C/ Nicol\'as Cabrera 13-15, Universidad Autónoma de Madrid, 28049 Madrid, Spain\\[2mm]
$^3$ Departamento de Física Teórica, Universidad Autónoma de Madrid, 28049 Madrid, Spain}

\small{\bf Abstract} \\[5mm]
\end{center}

\begin{center}
\begin{minipage}[h]{16.0cm}
  Type II theories can be described as the endpoint of closed string
  tachyon condensation in certain orbifolds of supercritical type 0
  theories. In this paper, we study solitons of this closed string
  tachyon and analyze the nature of the resulting defects in critical
  type II theories. The solitons are classified by the real K-theory
  groups KO of bundles associated to pairs of supercritical
  dimensions. For real codimension 4 and 8, corresponding to
  $KO(\IS^4)=\IZ$ and $KO(\IS^8)=\IZ$, the defects correspond to a
  gravitational instanton and a fundamental string, respectively.  We apply
  these ideas to reinterpret the worldsheet GLSM, regarded as a supercritical
  theory on the ambient toric space with closed tachyon condensation
  onto the CY hypersurface, and use it to describe charged solitons
  under discrete isometries. We also suggest the possible applications
  of supercritical strings to the physical interpretation of the
  matrix factorization description of F-theory on singular spaces.
\end{minipage}
\end{center}
\newpage
\setcounter{page}{1}
\pagestyle{plain}
\renewcommand{\thefootnote}{\arabic{footnote}}
\setcounter{footnote}{0}

\vspace*{1cm}

\setcounter{tocdepth}{2}

\tableofcontents

\vspace*{1cm}

\section{Introduction}

The understanding of open string configurations experienced a major breakthrough by the realization that they can be constructed as open
string tachyon solitons of suitable non-supersymmetric systems of
D-brane/antibrane pairs (see e.g. \cite{Sen:1998rg,Sen:1999mg}). This
formulation has clarified that K-theory is the right tool to classify
D-brane charges (and thus RR fields)
\cite{Witten:1998cd,Moore:1999gb}. Moreover, it has allowed
simplifying the analysis of involved supersymmetric D-brane
configurations, in terms of non-supersymmetric parent brane-antibrane
configurations (see e.g. \cite{Collinucci:2008pf,Collinucci:2014qfa}).

Recently, the study of closed string tachyons in the supercritical heterotic strings of  \cite{Hellerman:2004zm,Hellerman:2004qa,Hellerman:2006ff} 
has revealed a similar structure \cite{Garcia-Etxebarria:2014txa}, providing a closed tachyon soliton description for NS5-branes, and other objects. This motivates the possibility of exploiting closed tachyon solitons in other supercritical strings to describe objects beyond D-branes.

 In this paper we undertake this task for type II superstrings, realized as the endpoint of closed string tachyon condensation on certain orbifolds of supercritical type 0 theories \cite{Hellerman:2006ff}. We argue that the tachyon condensation picture describes annihilation/nucleation of pairs of bundles with $SO$ structure group, associated to supercritical dimensions, and use it to classify possible tachyon solitons in terms of real K-theory groups KO. We focus on the $\IZ$-valued objects, corresponding to real codimension 4 and 8 solitons. Our results are:
 
 $\bullet$ The codimension 4 soliton corresponds to a gravitational instanton in the critical type II theory. The tachyon condensation can be regarded as a surgery operation inserting a contribution to $\tr R^2$ at the core of the tachyon soliton. We provide several arguments supporting this interpretation, by using D-brane probes, and by arguing that the worldsheet CFT in the presence of the corresponding tachyon soliton corresponds to a $(4,4)$ singular CFT compatible with the right asymptotics in target space to describe $A_n$ singularities.

$\bullet$ The codimension 8 soliton in type IIA corresponds to a fundamental string (and by T-duality it corresponds to a momentum mode in type IIB). The F1 charge of the configuration is induced by the existence of a topological Chern-Simons coupling appearing at one-loop in the supercritical theory, as we explicitly compute. The coupling is a natural generalization of the type IIA $B_2\wedge X_8$ coupling, evaluated on the K-theory class of the supercritical bundles.

We describe several incarnations of these ideas in familiar
systems. We interpret the worldsheet GLSM as a supercritical theory
defined on an ambient toric space with a closed tachyon condensate
restricting the dynamics to the CY hypersurface. This interpretation
allows an interesting description of the discrete gauge symmetries
associated to discrete isometries of the CY (abelian or non-abelian),
and the construction of charged strings as closed tachyon solitons,
along the lines of \cite{Berasaluce-Gonzalez:2013sna}.

We finally present some suggestions to apply these results to the physical explanation of the matrix factorization description of F-theory in singular spaces \cite{Braun:2011zm,Braun:2012nk,Collinucci:2014taa}. Although the analysis is not complete, we present several suggestive hints, that we hope serve as stepping stone towards a full understanding in future works.

The paper is organized as follows. In Section \ref{supercritical-intro} we review the construction of orbifolds of supercritical type 0 theories  and their closed string tachyon condensation to type II theories (hence we dub the former `supercritical type II theories'). In Section  \ref{sec:general-view} we describe generalities of solitons in closed tachyon condensation, like their relation to K-theory (section \ref{sec:asymptotics}) and representatives using the Atiyah-Bott-Shapiro construction (section \ref{sec:abs}).

In Section \ref{sec:codim-four} we study the real codimension 4
soliton, and argue that it is a gravitational instanton. In section \ref{kk-codim4} we suggest this result from the structure of chiral zero modes in the core of the soliton from dimensional reduction of the bulk fields. In section \ref{sec:induced} we compute the effect of the soliton on D-brane probes, and show it corresponds to inducing lower-dimensional D-brane charges, exactly as for a $\tr R^2$ contribution. Finally in section \ref{sec:cft} we show the worldsheet theory in the soliton background is  a $(4,4)$ singular CFT compatible with the properties of the $\IC^2/\IZ_n$ CFT.

In Section \ref{sec:codim-eight} we study the real codimension 8
soliton, and argue that it is a fundamental string in type IIA (and a
momentum mode in IIB). The soliton charge arises from a novel
Chern-Simons coupling which arises at one-loop in supercritical IIA
theory, generalizing the 10d type IIA coupling $B_2X_8$. In section
\ref{sec:oneloopiia} we review the latter. In section
\ref{sec:anomaly-superiib} we compute the one-loop anomaly in the
supercritical type IIB theory, and in section \ref{sec:cssuperiia} we
relate it to the Chern-Simons coupling. In section
\ref{sec:nsbrane-anomaly} we take a brief detour to show that this
coupling precisely cancels the anomaly of the magnetically charged
NS-brane of supercritical IIA theory. In section \ref{sec:other} we describe other defects classified by $\IZ_2$-valued KO groups.

In Section \ref{sec:glsm} we interpret the familiar worldsheet GLSM as
a supercritical string closed string tachyon condensation, and use
this description to construct charged objects under discrete
isometries of the CY hypersurface. In Section \ref{sec:fth} we
describe applications to interpret physically the matrix factorization
description of F-theory on singular spaces. Section \ref{conclusions}
contains our final remarks.

Appendix \ref{appA} provides some details about the form of the ABS construction for bundles with SO structure group. Appendix \ref{app:caloron} describes the  construction of the codimension 4 defect with one transverse $\IS^1$ dimension, in terms of caloron solutions constructed as infinite periodic arrays of instantons.

\section{Review of supercritical strings and tachyon condensation}
\label{supercritical-intro}

In this section we review the construction of supercritical string theories flowing to 10d type II string theories upon a closed string tachyon condensation which eliminates the extra coordinates, following \cite{Hellerman:2006ff} (see also \cite{Hellerman:2004zm,Hellerman:2004qa}). 

The `supercritical IIA/B' string theories are constructed as $\IZ_2$ quotients of the supercritical extension of type 0A/B theories \footnote{Actually, there exist modular-invariant supercritical extension of the IIA/B GSO projection, but only for dimensions $10+8k$ \cite{Chamseddine:1991qu}. We will not discuss these supercritical theories.}. 

Supercritical type 0A/B theories defined in flat $10+n$ dimensions are described by the 2d real fields $(X^M, \psi^M)$ (forming $(1,1)$ multiplets) in the vector representation of $SO(10+n)$. Cancellation of the central charge requires including a timelike linear dilaton background $V^\mu$, satisfying
\begin{equation}
V^\mu V_\mu=-\frac{n}{4\alpha'}.
\label{dilaton-back}
\end{equation} There is a GSO projection by $(-1)^{F_w}$, where $F_w$ is total
worldsheet fermion number. The supercritical 0A (0B) theory
corresponds to choosing opposite (equal) GSO projections on the left-
and right-movers. There is an NSNS sector containing a real closed
tachyon scalar, and massless graviton, 2-form and dilaton fields, and
a RR sector containing massless $p$-form potentials (with $p$ odd/even
for the 0A/0B theories) coming in two sets (denoted `electric' and
`magnetic').

In order to connect with the 10d type II theories, we must consider the theories in $10+2k$ dimensions, namely with $n=2k$. We split the coordinate fields as the `critical' $X^\mu$, $\mu=0,\ldots, 9$, and the `supercritical' $x_m$, $y_m$, with $m=1,\ldots, k$. The supercritical IIA/B theories are defined as the orbifold by the action $g\equiv (-1)^{F_{L_w}}\cdot {\cal R}$, where $(-1)^{F_{L_w}}$ is left-moving worldsheet fermion number, and  ${\cal R}$ is a spacetime $\IZ_2$ action flipping the $k$ extra coordinates $y^m\to -y^m$, while leaving the others invariant $x^m\to x^m$. Note that for the 10d critical case ($k=0$), this is just the familiar orbifold turning the 10d 0A/B theory into the 10d IIA/B theory, see e.g. \cite{Dixon:1986iz,Seiberg:1986by,Bergman:1999km}. 

The  spacetime low-lying spectrum for these `supercritical type II theories' is: 

\smallskip

$\bullet$ The $g$-untwisted sector is the ($\IZ_2$ projected version of the)  NSNS and RR sectors, and describes $(10+2k)$-dimensional fields, namely a real tachyon $T$, a massless graviton, 2-form and dilaton, and RR $p$-form gauge potentials (with $p$ odd/even for the IIA/B).

$\bullet$ The $g$-twisted sectors are denoted by NS-R and R-NS sectors, and describe fields localized at the $(10+k)$-dimensional locus $y^m=0$.  The massless fields correspond in the NS-R sector to a $(10+k)$-dimensional  vector-spinor field $\psi^M_\alpha$, where $M$ runs through the $10+2k$ coordinates, and $\alpha$ denotes a bi-spinor of $SO(1,9+k)\times SO(k)$, with an overall chirality projection (i.e. a chiral spinor of $SO(1,9+2k)$ decomposed with respect to $SO(1,9+k)\times SO(k)$). The vector-spinor field can be decomposed in terms of gravitinos and Weyl spinors. The R-NS sector contains a similar set of fermions, with opposite/same overall chirality in the IIA/B theories.

These supercritical IIA/IIB theories are connected to the critical 10d IIA/IIB theories by condensation of the closed string tachyon in the NSNS sector. Following \cite{Hellerman:2006ff}, the tachyon couples as a worldsheet $(1,1)$ superpotential, 
\begin{equation}
\Delta {\cal L}\, =\, \frac{i}{2\pi}\int d\theta^+d\theta^-{\cal T}(X)\, =\; -\frac{1}{2\pi} \sqrt{\frac{\alpha'}{2}} F^M \partial_M {\cal T}(X) + 
\frac{i\alpha'}{4\pi} \partial_M\partial_N{\cal T}(X)\, {\tilde \psi}^M \psi^N
\label{wsupo}
\end{equation}
where $F^M$ is the auxiliary field for the superfield $X^M$. Upon use of its equations of motion $F^M\sim G^{MN}\partial_M {\cal T}$, we obtain  a worldsheet potential term
\begin{equation}
V\, =\, \frac{\alpha'}{16\pi} G^{MN}\partial_M{\cal T}\partial_N{\cal T}
\end{equation}

One can now study tachyon profiles which lead to dimension change within type 0 theories. This is achieved by pairing the extra coordinates, with mass terms
\begin{equation}
W={\cal T}=\mu \exp(\beta X^+) \itach_{mn} \,x_m \, y_n.
\label{type0-tach}
\end{equation}
where we take $\itach_{mn}$ to be a matrix of rank $k$ at generic points
of the critical dimensions, given in terms of the tachyon as
\begin{equation}
\itach_{mn}=\partial_{x_m}\partial_{y_n} {\cal T}
\label{tachyon-matrix}
\end{equation}
where in this last expression we ignore the dependence in $X^+$. This is because the light-like dependence on $X^+$ was introduced in \cite{Hellerman:2006nx,Hellerman:2006ff,Hellerman:2006hf,Hellerman:2007fc}\footnote{For other works on light-like tachyon condensation, see \cite{Hellerman:2007zz,Hellerman:2007ym,Hellerman:2008wp}.}  to make the system solvable in this case of constant $\itach$. In general, we will be interested in situations with non-constant $\itach$, for which the dynamics is not exactly solvable. However, we will focus on topological properties like tachyon solitons and their charges, so in what follows we drop this dependence in $X^+$.

The superpotential (\ref{type0-tach}) must be linear in the $y$'s due to the $\IZ_2$ symmetry (which is an R-symmetry in the worldsheet), while the linearity in the $x$'s can be regarded as the first term in a general Taylor expansion (and higher terms will be irrelevant in our discussion). 

The above tachyon condensation makes the coordinates $x$, $y$ disappear, and its endpoint is critical 10d type IIA/B theory \footnote{Actually, there is an abuse of language here (and throughout the paper), since the tachyon condensation process produces a light-like BPS configuration rather than the 10d vacuum of these theories.}. As shown in the references, the 2d dynamics ensures that the spacetime background is renormalized as required for consistency (in particular the dilaton (\ref{dilaton-back}) is redefined, and disappears in condensation to 10d). We will refer to the 10d locus $x=y=0$ where the type II theory lives as the ``critical slice''.

\section{Closed tachyons and solitons in type II}
\label{sec:general-view}

As done in \cite{Garcia-Etxebarria:2014txa} in the heterotic context, nontrivial tachyon profiles in type 0 or type II theories can be associated to different kinds of solitons. In this section we develop this connection in some detail.

\subsection{Closed tachyons, asymptotics, and K-theory}
\label{sec:asymptotics}

Consider the case of non-constant mass matrix \eqref{tachyon-matrix}. The basic intuition is that in regions where the rank of $M$ is $k$, tachyon condensation proceeds as above and eliminates the supercritical coordinates. On the other hand, there is interesting new physics at loci where its rank lowers, i.e. some combinations of coordinates become massless at some locus, because of zero eigenvalues of the tachyon matrix. 

In this study, we may write down particular tachyon profiles with
zeroes, but generically they will not satisfy the spacetime equations
of motion and so the tachyon will not describe a marginal
perturbation. 
It is then difficult to argue that the tachyon zeroes will
survive the condensation, since this requires knowledge of the RG
flow. A simple way around this difficulty is to consider cases in
which the tachyon matrix zeroes are topologically protected. This is
the case when the tachyon profile is topologically non-trivial,
namely, when the fiberings of the supercritical dimensions $x$ and $y$
over the critical slice are not isomorphic.  For instance, a section
of a bundle with nontrivial Euler class must always vanish at its
Poincar\'{e} dual locus.

In other words, the tachyon is sensitive to the difference bundle (the
K-theory class) of the bundles described by the fiberings of $x$ and
$y$ over the critical slice.\footnote{\label{f4}In many of our examples the
  critical slice is $\bR^n$, so strictly speaking there are no
  non-trivial bundles. As we discuss momentarily, we always impose
  finite energy at infinity for our bundles, so effectively we have in
  mind bundles on the compactification of flat space.} We will see
shortly that the physical meaning of this K-theory class is that
supercritical dimensions can be nucleated/annihilated by pairs via
closed string (constant) tachyon condensation.

The intuition is very close to that of open string tachyons in brane antibrane annihilation \cite{Witten:1998cd}. Indeed, as in the open string case, the loci where the tachyon matrix decreases its rank corresponds to the presence of extra degrees of freedom associated to a (closed) tachyon soliton\footnote{This was explored in the  heterotic context in \cite{Garcia-Etxebarria:2014txa}, see also \cite{Berasaluce-Gonzalez:2013sna}.}. These solitons are therefore naturally classified by the real KO-theory groups.

An essential point of this construction is that tachyon profiles
describing solitons are in one-to-one correspondence with K-theory
classes of the $x$, $y$ bundles. Indeed,  as mentioned above the
tachyon matrix $\itach$ is a bundle isomorphism from $x$ to $y$, which
specifies a $K$-theory class uniquely as follows: Given a tachyon $T$ over some compact\footnote{Recall footnote \ref{f4}.} base space $X$, one may construct a complex vector bundle $E$ by finding an open cover $\{U_j\}$ of $X$ such that the tachyon has full rank on the overlaps $U_i\cap U_j$ (for instance one in which one of the $U_i$ is a tubular neighbourhood of the $\det T=0$ locus) and setting the transition functions to be just $T$. From this we can further construct a $K$-theory class representative as $(E,1)$ where $1$ is a trivial vector bundle.

 However, one must show that this mathematical correspondence has some physical meaning, for instance showing that the tachyon profile requires the $x$, $y$ bundles to form a representative of the tachyon K-theory class. Physically, the equivalence between topologically nontrivial tachyon profiles and the supercritical $x$, $y$ bundles arises from two assumptions about the theory:\begin{itemize}
\item Tachyons describing a localized soliton must relax to the vacuum far away from the core. In practical terms this means that $\mathcal{T}\rightarrow \mathcal{T}_0$ as $r\rightarrow\infty$, where $\mathcal{T}_0$ is the vacuum expectation value of the tachyon.
\item Configurations of finite action on $\IR^k$ can be extended to differentiable configurations on its one-point compactification $\IS^k$, modulo gauge transformations. This is an assumption about the form of the action near the vacuum. For instance, in Yang-Mills theory in $k$ dimensions, one must have 
\begin{equation}F_{\mu\nu}\rightarrow0\end{equation}
faster than $r^{-k/2}$ for the action to be finite. Since this implies that the connection is flat at infinity, the configuration can be extended to $\IS^k$, possibly in a singular gauge. Similarly, assuming that near the tachyon vev the action takes the form
\begin{equation}S=\int \vert\mathcal{D}\mathcal{T}\vert^2-V(\mathcal{T})\end{equation}
for some effective potential $V(\mathcal{T})$ with a minimum around $\mathcal{T}_0$, finite action means that $\mathcal{D}\mathcal{T}\rightarrow0$ faster than $r^{-k/2}$.
\end{itemize}

We should remark that whereas the first assumption is quite reasonable
and could be considered as part the definition of a soliton, the
second is a strong statement about the action in a regime we do not
really control. The supercritical description is valid for small vevs
of $\mathcal{T}$ and even though it is possible to ascertain the
endpoint of the condensation for several different profiles, we know
nothing about the effective action there. In fact, this phenomenon is
not new. The same happens in the context of brane-antibrane
annihilation in type IIB, where the tachyon profiles which yield
$D$-branes asymptote to the vacuum expectation value of the
tachyon. Although some is known about the effective action of the
tachyon in this case (see \cite{Sen:2004nf} for a review), it is
difficult to argue in a controlled way that the covariant derivative
of the IIB tachyon should vanish at infinity to satisfy finite energy
conditions.

In any case, once we assume that $\mathcal{D}\mathcal{T}\rightarrow0$,
we obtain a link between the tachyon and the supercritical bundles. We will review the argument in the more familiar case of D9-$\overline{\text{D}9}$ tachyon condensation in type IIB this more familiar case first, and then briefly discuss the the closed-string version of the phenomenon.

In type IIB in the presence of extra D9-$\overline{\text{D}9}$ pairs, there is a tachyon $\mathcal{T}_{IIB}$ in the bifundamental of the gauge group. A tachyon vev $\vert\mathcal{T}_{IIB}\vert\rightarrow\infty$ corresponds to complete annihilation to the type IIB vacuum. Different tachyon profiles describe different configurations in the critical theory after tachyon condensation. 

The tachyon asymptotics impose constraints on both D9 and
$\overline{\text{D}9}$ connections. For concreteness, consider a single brane-antibrane pair and a tachyon profile of the
form $\mathcal{T}_{IIB}=z$. This is well-known to describe a D7-brane
sitting at $z=0$ after condensation. Far from $z=0$ the tachyon is
non-vanishing, indicating complete annihilation. However, there is one
extra requirement to be imposed. As discussed above, one should have
$\mathcal{D}_\mu\mathcal{T}\rightarrow 0$ sufficiently fast. Plugging
the tachyon profile we have chosen, this means that
\begin{equation}
\partial_z \mathcal{T}-(A_{D9}-A_{\overline{D9}})\mathcal{T}\rightarrow 0
\label{esas}
\end{equation}
or, in other words, that asymptotically

\begin{equation}A_{D9}-A_{\overline{D9}}\rightarrow \frac{1}{z}.
\label{D91}
\end{equation}
One may measure the first Chern class of the K-theory class $(E,F)$, where $E$ is the D9 gauge bundle and $F$ the $\overline{\text{D}9}$ gauge bundle, by integrating the Chern-Simons form of the connection \eqref{D91}. In this case, one obtains an induced D7-brane charge of 1, as befits a D7-brane after condensation. We thus see that in order to be able to draw the correspondence between a tachyon profile and a particular configuration after condensation it is essential to specify the behaviour at infinity. 

We will assume that an analogous discussion holds in the supercritical
theory. The projection of the covariant derivative $\nabla_\mu \itach$ to the
critical slice is analogous to the covariant derivative of the tachyon
in the IIB example above. We will hypothesize that
$\nabla_\mu \itach\rightarrow 0$ plus terms of order $r^{k/2}$ as we
near infinity for tachyon profiles describing solitons of codimension
$k$. By expanding the covariant derivative as in \eqref{esas}, we get
a sum of Christoffel symbols which act as connection coefficients over
the $x,y$ bundles. The same arguments as above show that a particular
tachyon matrix specifies a pair of $x,y$ bundles.

In either case, tachyon solitons are classified by K-theory classes of the corresponding bundles. As usual, we are interested in compactly supported K-theory classes in $\IR^p$, which correspond to the K-theory groups of $\IS^p$.

\subsection{The ABS construction}
\label{sec:abs}

We will soon turn to the construction of particular interesting
tachyon solitons. To construct a soliton of real codimension $k$, the
matrix $\itach$ is a map from $\IR^k$ to the difference $SO(n)$ (which
is the subgroup broken by the tachyon background).  These topological
classes of such maps are classified by the behaviour of the map at the
$(k-1)$-dimensional sphere at infinity, i.e. the homotopy group
$\Pi_{k-1}(SO)$. As we have the freedom of introducing extra pairs of
supercritical dimensions, the rank of the bundles remains
undetermined, more precisely we take it in the stable range. These
stable homotopy groups are in correspondence with the K-theory classes
of pairs of bundles with $SO$ structure group. Namely, the topological
classes of bundles, defined modulo creation/annihilation of pairs of
bundles (physically realized as introduction/removal of topologically
isomorphic supercritical dimensions.)

In the $U(n)$ case, a simple representative of the non-trivial
homotopy classes is obtained by the Atiyah-Bott-Shapiro (ABS)
construction \cite{Atiyah:1964zz}, which has been exploited in the
construction of open tachyon solitons as well
\cite{Witten:1998cd}. For the case of complex bundles, the K-theory
group is $K(\IS^{k-1})=\IZ$ for even $k$, $k=2p$. The two bundles are
taken to describe the chiral spinor bundles $S^{\pm}$ over $\IR^k$,
namely $U(n)$ bundles with $n=2^{p-1}$, and the tachyon background
generating the K-theory group is
\begin{equation}
  \itach= \Gamma \cdot \vec{x}
\label{abs}
\end{equation}
where $\Gamma$ denote the Dirac matrices in Weyl representation, so we indeed have a map $M:S^+\to S^-$. The $\IZ$-valued charge can be identified with the non-trivial Chern classes $\tr F^{p}$, $k=2p$ along $\IR^k$.

The $SO$ version of the above construction is obtained by embedding $U(n)\subset SO(2n)$, and (for the physically relevant case of 10 dimensional string theory) describes topologically non-trivial configurations only for $KO(\IS^{k-1})=\IZ$ for $k=4,8$ and $KO(\IS^{k-1})=\IZ_2$ for $k=1,2,9,10$.
We will mostly focus on the $k=4,8$ cases, whose $\IZ$-valued charge can be realized in terms of the Pontryagin classes $\tr R^2$, $\tr R^4$ of the difference bundle over $\IS^4$, $\IS^8$, respectively.  Explicit expressions for the real tachyons in the $k=4,8$ cases can be found in appendix \ref{appA}.

Notice that the above discussion only uses the supercritical tachyon
$\mathcal{T}$ and the supercritical bundles, hence the tachyon
condensation and its topological solitons work the same way both for
type IIA and IIB string theory. This already provides information
about the 10d solitons which can be obtained using different tachyons
of the form \eq{abs}: They cannot be D-branes, for a T-duality in the
10d theory along a worldvolume direction would change the codimension
of the soliton, whereas the same T-duality before tachyon condensation
merely turns 0A to 0B leaving the tachyon profile \eq{abs} and thus
the codimension of the objects untouched.

So focusing e.g. on the $\IZ$-valued cases, we are looking for
e.g. codimension 4,8 objects in type II theory which upon T-duality
along their worldvolume turn into objects of the same dimension. For
codimension 4, the natural candidates are gravitational instantons or NS5-branes. For codimension 8, these would be fundamental strings.

 In the following sections we consider these two cases in more detail, and describe the nature of the corresponding soliton in the 10d spacetime. 

\section{Codimension 4 solitons are gravitational instantons}
\label{sec:codim-four}

Let us consider the tachyon background associated to $\Pi_3(SO)=\IZ$, which describes  supercritical bundles with non-trivial first Pontryagin class. For concreteness, we consider $4+4$ supercritical dimensions, and fiber them  such that the $SO(4)_x \times SO(4)_y$ bundles have non-trivial $\tr R_x ^{\, 2}-\tr R_y^{\,2}$ over an $\IR^4$ in the 10d critical slice.

Using e.g. the profile (\ref{abs}) as prototype, the tachyon
condensation removes the supercritical dimensions everywhere except at
the origin in $\IR^4$. Just like for open string tachyon solitons, we
expect this to signal the presence of a left-over real codimension-4
topological defect. One may be tempted to propose that this is an
NS5-brane, as occurred in the supercritical heterotic case
\cite{Garcia-Etxebarria:2014txa}. In the following we rather show that
it corresponds to a gravitational soliton, carrying the gravitational
charges of $\bC^2/\bZ_2$, namely, a localized contribution to
$\tr R^2$, where here $R$ is the curvature 2-form of the {\em tangent}
bundle in $\IR^4$.

The tachyon and its condensation proceeds essentially in the same way
in both type IIA and type IIB theories. We focus on the latter case,
since the nature of the endpoint and the corresponding charges are
easier to identify, essentially because in type IIB theory, ADE
geometries localize chiral field content at the singularities, whereas
for type IIA theory the localized modes inherit the non-chiral nature
of the ambient 10d theory. The result of type IIB theory can be
translated to type IIA theory by a T-duality along the worldvolume
dimensions of the defect (even in the supercritical setup).

In addition, in setups with one compact $\IS^1$ dimension transverse to the defect it is possible to T-dualize the gravitational instantons into NS5-branes. The corresponding configurations and relevant technical details are briefly discussed in appendix~\ref{app:caloron}. This hence provides a realization of the NS5-brane as a closed tachyon soliton.

\subsection{Chiral worldvolume content}
\label{kk-codim4}

Although the tachyon condensation process is highly non-trivial, one
could hope to compute the worldvolume chiral spectrum by simple
dimensional reduction of the 10d fields using the coupling to the
physical tachyon background, as done in
\cite{Hellerman:2004zm}. However, although our profile \eq{abs}
specifies the right K-theory class of the soliton, and it is expected
to flow to the physical tachyon background, as it stands it does not
solve the spacetime equations of motion.  Therefore, even if we focus
on the chiral sector, the computation of the spectrum is not safe
against leaking of modes off to infinity in the supercritical
dimension, as we now describe.

If we attempt the dimensional reduction of the fermions using the
tachyon profile \eq{abs}, the following picture emerges. For each
fermion, which in the supercritical theory live in the
$10+k$-dimensional orbifold fixed locus (we remind the reader that
$k=4$ in this section), one can construct two different sets of zero
modes. One of them is localized along a $10$-dimensional slice which
will become the critical type II spacetime after tachyon
condensation. Presumably these describe the 10d fermion content of the
theory, as happens for the heterotic in \cite{Hellerman:2004qa}. There
is a second set, which is instead localized along a different
$10$-dimensional slice, which intersects the critical slice on the
worldvolume of the defect. This different set of fermions is
associated with the $X^\mu=0$ branch of the superpotential obtained
from the tachyon matrix \eq{abs}; they may be thought of as a
supercritical version of the localized worldvolume fermions of the
gravitational instanton, before condensation. From the point of view
of the critical slice, they are 6d localized fermions, even though
they propagate in 4 extra supercritical dimensions. It is expected
that the physical tachyon localizes the modes in the latter, leading
to genuinely 6d chiral modes; this phenomenon is however not robust,
as such modes can leak off to infinity along the 4 supercritical
dimensions. From a different but related perspective, the computation
of the chiral spectrum using the index theorem is ambiguous due to
contributions from the $\eta$-invariant at infinity.

This picture suggests that localized fermions indeed arise from
tachyon condensation with the superpotential \eq{abs}, but as stressed
above, we cannot trust this tachyon profile in spacetime. Fortunately,
other arguments, which we now proceed to review, show clearly that the
endpoint of the condensation is a gravitational instanton.

\medskip

\subsection{Induced D-brane charges}
\label{sec:induced}

The problem of tachyon condensation in supercritical type 0 theories with a codimension 4 ABS profile is in principle similar to what happened in the supercritical heterotic tachyon solitons \cite{Garcia-Etxebarria:2014txa}, with non-trivial first Pontryagin class in the supercritical dimensions. There, the localized defect  was argued to be an NS5-brane, which actually carries the same charges as $\tr R^2$. What allowed to read the charges of the soliton more easily there was the existence of couplings such as  $B_6\, \tr R^2$ (where $B_6$ is the potential dual to the 2-form), which survived in the supercritical setup and included contributions such as $B_{6+n}\tr R_N^2$, where $R_N$ is the curvature of the normal bundle the critical slice. However, a similar argument does not apply straightforwardly to type 0 theories, since there are no topological couplings of this kind. 

We can however use a different strategy to detect the physical charges
carried out by supercritical dimension bundles with non-trivial first
Pontryagin class. The key idea is to introduce D-brane probes, on
which the topological charge $\tr R^2$ will induce lower-dimensional
D-brane charges.

D-branes in type 0 theories are discussed in
e.g. \cite{Bergman:1997rf,Bergman:1999km}; most of these results apply
to the supercritical theories in a straightforward manner. Basically,
there is one D$p$-brane per RR gauge potential (hence, two D-branes,
dubbed `electric' and `magnetic' for each allowed value of $p$). On a
D$p$-brane stack of fixed kind, the spectrum is given by gauge bosons
and scalars associated to the transverse dimensions. For overlapping
electric and magnetic D$p$-branes, the mixed sector gives rise to a
Majorana-Weyl fermion of the rotation group of the full
$(10+2n)$-dimensional space (we assume $n\in 4\bZ$ here), suitably
decomposed with respect to the worldvolume Poincar\'e group.

We now consider D-branes in the $\IZ_2$ quotient describing the supercritical type II configuration. Specifically, we will consider only supercritical D-branes which are localized on a submanifold contained in the 10d critical slice, and which span all supercritical directions, including the orbifolded ones. In this way, tachyon condensation takes place along the worldvolume of the D-brane and we avoid the tricky question of determining what happens to other RR charges which exist in the supercritical theory but not in the type II endpoint.

 Since the supercritical type II theories are constructed as $\IZ_2$ orbifolds of supercritical type 0 theories, the D-branes of supercritical type II arise in the parent type 0 theory $\IZ_2$ as invariant pairs of D-branes \footnote{In fact, this is completely analogous to what happens for critical 10d type II theories, constructed as $\IZ_2$ orbifolds of type 0 theories \cite{Bergman:1997rf,Bergman:1999km}.}.  Since the $\IZ_2$ exchanges the electric and magnetic D-branes of the parent type 0 theory, the supercritical type II D-brane is a coincident pair of electric and magnetic supercritical type 0 D-branes of the corresponding dimension. Note that we have to include electric and magnetic branes to guarantee invariance under the $\IZ_2$ at the orbifold fixed locus; if the brane were away from it, it would be completely consistent to have a single electric or magnetic brane.

For concreteness, we focus on type 0B theory, in $10+4+4$ dimensions, and consider a pair of electric and magnetic D13-branes (denoted $e$ and $m$ in the following) along the directions $x^0$, $x^1$ and the $\IR^4$ parametrized by $(x^6,\ldots, x^9)$ in the 10d critical slice, and extending over the 4 orbifold-even $x$ coordinates as well as the 4 orbifold-odd $y$ coordinates. We are interested in computing the worlvolume Chern-Simons couplings to the curvatures, in order to read out the D-brane charges induced by the curvature of the normal supercritical dimensions. Such couplings are intimately related to the worldvolume fermion content, as required by anomaly inflow arguments \cite{Green:1996dd,Cheung:1997az}, so we determine the latter. As reviewed above, the D13$_e$-D13$_e$ and D13$_m$-D13$_m$ sectors do not contain fermions, while the mixed D13$_e$-D13$_m$ (and  D13$_m$-D13$_e$) give rise to two fermions (one per sector) of fixed chirality with respect to the total $SO(1,9+4+4)$ group, decomposed under the $SO(1,5)\times SO(4)_{2345}\times SO(4)_x\times SO(4)_y$ group. This worldvolume spectrum is non-chiral in $14$ dimensions but at the orbifold fixed locus only one linear combination of the fermions survives, leading to a localized contribution to the  10d anomaly. 

To be more precise, the theory has two 14d fermions, of opposite 14d chirality. They can be decomposed in terms of 10d and 4d spinors as $\ket{C}_{10}\ket{C}_4+\ket{S}_{10}\ket{S}_4$ and $\ket{C}_{10}\ket{S}_4+\ket{S}_{10}\ket{C}_4$, where $\ket{C}$ and $\ket{S}$ denote spinors of opposite chiralities. The $\IZ_2$ orbifold flips the sign of the second 14d spinor, which implies that one cannot regard the $\IZ_2$ orbifold as having a simple geometric action on the 4d spinors (this would swap the sign of one of the two 4d spinors). Hence, the computation of the localized anomaly at the  $y=0$ locus is subtle, but can be dealt with as follows. In fact, the action of the orbifold on the chiral fermions is a single copy of its action on the bispinor generating the bulk RR forms. Later, in section \ref{sec:codim-eight}, we will see that both anomaly cancellation in 0B theory and reduction to the anomaly polynomial in IIB after condensation requires a specific form for the anomaly polynomial of the RR forms.

  Inspired by the form of the RR anomaly polynomial, a natural guess for the anomaly polynomial of the chiral fermions turns out to be
\begin{align}P_{10}=\frac{\hat{A}(R^{D13\cap \{y=0\}}_T)}{\hat{A}(R^{D13\cap \{y=0\}}_{N})}\,e(R_y),\end{align}
where $R^{D13\cap \{y=0\}}_T$ is the curvature of the tangent bundle to the intersection of the $y=0$ locus and D13-brane worldvolume, and $R^{D13\cap \{y=0\}}_N$ its normal bundle. The last factor, $e(R_y)$, is the Euler class of the $y$-bundle, which will localize anomalies in a $6$-dimensional locus. This will become the D5-brane after tachyon condensation (see section \ref{sec:codim-eight}).

One can now apply standard anomaly inflow arguments \cite{Green:1996dd,Cheung:1997az}, considering two branes intersecting along the critical slice, to describe the Chern-Simons coupling on the D13-brane worldvolume in terms of curvatures of the corresponding bundles. We denote by $R$, $R_N$, $R_x$, $R_y$ the curvatures of the tangent bundle along the D13-brane, and along the normal bundles in the directions 2345 and the supercritical $x$ and $y$ directions, respectively. Since the $\hat{A}$-genus is a multiplicative class, we have 
\begin{equation}
\hat{A}(R_T^{D13\cap \{y=0\}})=\hat{A}(R)\hat{A}(R_x) \quad , \quad \hat{A}(R^{D13\cap \{y=0\}}_N)=\hat{A}(R_n)\hat{A}(R_y).
\end{equation}
The resulting Chern-Simons couplings are
\begin{equation}
S_{\rm CS}\, =\, \int_{\rm D5} \sum_p C_p \wedge {\rm ch} (F)\, \Bigg(\, \frac{\hat{A}(R)}{\hat{A}(R_N)}\, \frac{\hat{A}(R_x)}{\hat{A}(R_y)} \Bigg)^{\frac 12}
\label{Dcs}
\end{equation}
(The Euler class contribution is rewritten here as the localization to
the D5 slice on the critical directions.) Notice that the $x$ and $y$
bundles appear asymmetrically in this expression. This difference between the $x$ and $y$ bundles is crucial for the
$K$-theory picture advocated throughout this paper to make
sense. Namely, isomorphic bundles in the $x$ and $y$ dimensions should
not generate new charges in the system, and indeed the contribution of
isomorphic $x$ and $y$ bundles to induced D-brane charges cancel in
(\ref{Dcs}); see below for an explicit example of this. Similar
expressions can be obtained for other D$p$-branes. Note that in the
critical case this reduces to the familiar expression in
\cite{Minasian:1997mm}.

Consider now a D13-brane in flat space, $R=R_N=0$, but in the presence of  non-trivial bundles in the supercritical dimensions. Expanding the above we have 
\begin{equation}
\frac{1}{48}\int_{\rm D13} C_{10}\wedge (\tr R_y^{\, 2}-\tr R_x^{\,2}).
\end{equation}
This implies that in the presence of supercritical bundles with non-trivial first Pontryagin class, there is an induced (supercritical) D9-brane charge on the volume of the D13-brane. Upon tachyon condensation, the extra dimensions disappear everywhere except at the core of the tachyon soliton, while each brane loses 8 worldvolume directions along the $x,y$ directions\footnote{This follows from applying the worldsheet arguments  in section \ref{supercritical-intro} to the open string sector describing the D-branes, i.e. the 2d fields associated to the $x,y$ directions are very massive and can be integrated out.}. The left-over defect in the endpoint critical type II theory must be such that it reproduces the induced D1-brane charge on a D5-brane probe. This nicely fits with the interpretation of the soliton as a localized defect supporting a non-trivial $\tr R^2$ along the tangent directions of critical type II theory.

\medskip

\subsection{Worldsheet CFT}
\label{sec:cft}
 
The above arguments show that the endpoint of tachyon condensation is
the critical type II theory with a localized, real codimension-4
object characterized by a nonvanishing $\tr R^2$, i.e. a gravitational
instanton. We will now argue that the worldsheet theory with the ABS
tachyon has the right properties for reproducing, after tachyon
condensation, the singular worldsheet CFT for a $\bC^2/\bZ_2$
singularity.

\medskip

The first piece of the argument concerns the topological charge of the
gravitational instanton that we just found. As we describe in
appendix~\ref{app:realification} the supercritical tachyon $\itach$ is
obtained by embedding the $SU(2)$ instanton into $SO(4)$. The induced
brane charge is given by the Pontryagin class of the real bundle,
which is defined to be the second Chern class of its complexification,
so the real embedding of the ABS instanton induces two units of
topological charge, compatible with a $\bC^2/\bZ_2$ singularity.

\medskip

A detailed analysis of the supercritical CFT gives further support for
this identification, as follows. As usual, take $X^\mu$ to be the
critical slice coordinates, $y^n$ the four supercritical orbifold-odd
coordinates, and $x^m$ the supercritical orbifold-even coordinates. As
discussed above, the matrix $\itach$ (\ref{tachyon-matrix}) of tachyon
derivatives with respect to the supercritical coordinates may be
regarded a mapping between the orbifold-odd and orbifold-even
supercritical tangent bundles. If we regard these as nontrivial
bundles over the critical slice parametrized by the $X^\mu$, the
tachyon describes a K-theory class over the critical slice, as
described in section \ref{sec:general-view}.
In particular, the closed string tachyon
\eqref{eq:supercritical-tachyon} (see appendix \ref{appA})
\begin{align} 
  \tach = X^\tau (\ov\eta_\tau)_{ab} x^a y^b = \frac{1}{2}\tr(\widetilde{\sx} \sX \sy)
  \label{ABS}
\end{align}
describes the K-theory class of a gravitational instanton. We have
argued for this interpretation using spacetime considerations. In the
following we will study the effects of this tachyon superpotential as
a worldsheet perturbation.

The equations of motion coming from~\eqref{ABS} (which, recall,
couples to the worldsheet as a superpotential) give a branch
structure. Consider for example $y\neq 0$. By a combined $SO(4)^3$
rotation we can put $y=(y_0,0,0,0)$ with $y_0\neq 0$ without changing
$\tach$. Now consider the equation of motion for $\widetilde{x}_\mu$,
which is
\begin{equation}
  \frac{\partial \tach}{\partial \widetilde{x}_\mu} =
  \frac{1}{2}\tr(\sigma^\mu \sigma^\tau)\, y_0 X^\tau = \pm y_0 X^\mu = 0
\end{equation}
for all $\mu$. (The sign is determined by
$\frac{1}{2}\tr(\sigma^\mu\sigma^\tau)=\diag(1,-1,-1,-1)$.) Since
$y_0\neq 0$, we have $X^\mu=0$, and from the $X$ equation of motion
also $\widetilde{x}^\mu=0$. A completely analogous argument holds when
$X$ or $\widetilde{x}$ are non-vanishing, so at the classical level we
find a three-branch structure, with each branch being determined by
which of $X,\widetilde{x},y$ is non-vanishing, and the other two
fields vanishing on that branch. The three branches meet at
$X=x=y=0$. This situation is very similar to that analyzed in
\cite{Witten:1994tz}, and for the same reasons as those given in
there, we expect the branch structure to disappear quantum
mechanically (we will review the details of the argument
momentarily). This can occur in the full CFT once we include quantum
corrections: the sigma model metric is corrected, and the branch
intersection is pushed to infinite distance down an infinite throat
developing at $X=0$. Such an infinite throat behavior close to $X=0$
is suggestive of light degrees of freedom (coming from wrapped branes)
living there, again indicating a singularity in the target
description. The real ABS instanton \eqref{ABS} is the minimal
instanton we can construct starting with a $SU(2)$ instanton, so it is
perhaps not surprising that the resulting configuration is the minimal
CY two-fold singularity.

\medskip

Finally, we will show that the sigma model given by \eq{ABS} has
$(4,4)$ worldsheet supersymmetry, as befits type II on
$\bC^2/\bZ_2$. Actually, proving $(0,4)$ will be enough; since the
arguments will be completely left-right symmetric, this will also
prove $(4,0)$. The fields come in supermultiplets of $(4,4)$
containing four bosons and four fermions of each chirality; we only
have to show that the interactions preserve this supersymmetry. A
$(1,1)$ superfield
\begin{align}
X\equiv X+\psi_+\theta_++\psi_-\theta_-+F\theta_-\theta_+
\end{align}
can be decomposed as two $(0,1)$ superfields
\begin{align}
X\equiv (X+\psi_+\theta_+)+\theta_-\lambda_-,\quad \lambda_-\equiv\psi_-+F\theta_+.
\end{align}
Then one can compute the $(0,1)$ superpotential from the $(1,1)$ superpotential \eq{ABS}  by integrating over $\theta_-$, which yields:
\begin{align}W_{(0,4)}=\int d\theta_- \tach=X_\mu\overline{\eta}^\mu_{mn} (\lambda^x)^m y^n+(\lambda^X)_\mu\overline{\eta}^\mu_{mn} x^m y^n+X_\mu\overline{\eta}^\mu_{mn} x^m (\lambda_y^n).\label{lag1}\end{align}
To show that the theory with superpotential $W_{(0,4)}$ actually has $(0,4)$ supersymmetry, it is enough to show that the ADHM equations are satisfied \cite{Witten:1994tz}. Following \cite{Garcia-Etxebarria:2014txa,Witten:1994tz}, it is convenient to write down the indices of the above superfields in terms of the would-be $SU(2)\times SU(2)$ $R$-symmetry. Uppercase unprimed indices will transform as a doublet over the first factor, whereas uppercase primed indices will transform in the same way under the second. We will choose the $R$-symmetry to act on the scalar superfields in such a way that $X$ transforms as $X^{AB}$, $x$ transforms as $x^{YY'}$, and $y$ transforms as $y^{A'B'}$. The three fields transform differently under $R$-symmetries\footnote{There is another scalar multiplet of $(0,4)$ not considered in \cite{Witten:1994tz}, namely $x^{YY'}$. The supersymmetry transformations are $\delta_\eta x^{YY'}=i\epsilon_{X'B'}\eta^{XX'}\psi_-^{B'Y'},\quad \delta_\eta\psi_-^{B'Y'}=\epsilon_{YB}\eta^{BB'}\partial_- X^{YY'}$. }, and this choice will ensure the $(0,4)$ symmetry of the theory. We still have to choose how the Fermi superfields transform under the $R$-symmetry; there is no need to be consistent with the rules we set up above for the scalar superfields\footnote{Indeed, doing so would ruin $(0,4)$ supersymmetry.}. We will take 
\begin{align}\lambda^X_{A'B'},\quad \lambda^x_{Y'Y}, \quad \lambda^y_{AB}.\end{align}

With this choice in mind, one may now use the identities 
\begin{align}2\overline{\eta}^\mu_{mn}&=\text{Tr}[\epsilon \overline{\sigma}^m\epsilon\sigma^\mu\epsilon\sigma^n]=\overline{\sigma}^m_{E'A}\epsilon^{AB}\sigma^\mu_{BC}\epsilon^{CC'}\sigma^n_{C'D'}\epsilon^{D'E'}\nonumber\\&=-\overline{\sigma}^m_{CD'}\epsilon^{D'A'}\sigma^\mu_{A'B'}\epsilon^{B'E'}\sigma^n_{E'C'}\epsilon^{CC'}=-\overline{\sigma}^m_{EC'}\epsilon^{CC'}\sigma^\mu_{CA}\epsilon^{AB}\sigma^n_{BD}\epsilon^{DE},\end{align}
to rewrite the superpotential \eq{lag1} as
\begin{align}W_{(0,4)}&=\frac12((\lambda^x)^m\overline{\sigma}^m_{E'A})\epsilon^{AB}(X^\mu\sigma^\mu_{BC})\epsilon^{CC'}(y^n\sigma^n_{C'D'})\epsilon^{D'E'}\nonumber\\&-\frac12(x^m\overline{\sigma}^m_{CD'})\epsilon^{D'A'}((\lambda^X)^\mu\sigma^\mu_{A'B'})\epsilon^{B'E'}(y^n\sigma^n_{E'C'})\epsilon^{CC'}\nonumber\\&-\frac12(x^m\overline{\sigma}^m_{EC'})\epsilon^{CC'}(X^\mu\sigma^\mu_{CA})\epsilon^{AB}((\lambda^y)^n\sigma^n_{BD})\epsilon^{DE}\nonumber\\&=\frac12\epsilon^{CC'}\left[\lambda^x_{E'A} X_C^Ay_{C'}^{E'}-\lambda^X_{A'B'}x_{C}^{A'}y^{B'}_{C'}-\lambda^y_{BD}X_C^Bx^{D}_{C'}\right].\end{align}

Defining $c^{CC'}=\frac12\epsilon^{CC'}$, the above may be rewritten as 
\begin{align}W_{(0,4)}&=c^{CC'}\left(\sum_{a=X,x,y} C^a_{CC'} \lambda^a\right),\nonumber\\C_{CC'}^x&= X_C^Ay_{C'}^{D'},\quad C_{CC'}^X= -x_C^{A'}y_{C'}^{B'},\quad C_{CC'}^y=- X_C^Bx^{D}_{C'}.\label{potwit}\end{align}

Showing that the theory has $(0,4)$ supersymmetry amounts to showing that the $C$'s satisfy the ADHM equations, explicitly given by (2.13) and (2.20) of \cite{Witten:1994tz}, which is easy. The first constraints are
\begin{align}
\frac{\partial C^a_{AA'}}{\partial X^{BY}}+ \frac{\partial C^a_{BA'}}{\partial X^{AY}}=\frac{\partial C^a_{AA'}}{\partial x^{YY'}}+ \frac{\partial C^a_{YA'}}{\partial x^{AY'}}=\frac{\partial C^a_{AA'}}{\partial y^{B'Y'}}+ \frac{\partial C^a_{AB'}}{\partial y^{A'Y'}}=0
\end{align}
where $a$ runs over $X,x,y$ and we have omitted the upper indices. The three different $C$'s above satisfy the equations. Secondly, we have to show that the constraint
\begin{align}
\sum_a C^a_{AA'}C^a_{BB'}+C^a_{BA'}C^a_{AB'}=0
\label{scalarcond}
\end{align}
is satisfied. Even though \eq{scalarcond} involves a sum over the three different sets of fermions, for \eq{potwit} the sum is identically zero for each fermion. This proves that the lagrangian under consideration actually has $(0,4)$ supersymmetry. In the same way we may prove that it has $(4,0)$, and since left and right-handed supercharges anticommute this shows $(4,4)$ supersymmetry.

We can now argue as in \cite{Witten:1995zh}, and follow what happens
to the right-handed $SU(2)\times SU(2)$ $R$-symmetry as we flow to the
IR (equivalently, as tachyon condensation takes place). The endpoint
of the condensation must be a $(4,4)$ nonlinear sigma model, and one
of the two $R$-symmetry $SU(2)$'s must be included as part of the
superconformal algebra. If we focus on the branch on which $X$ is
nonzero, the $SU(2)$ factor which acts nontrivially on $X$ cannot be
conserved and thus it must be the factor acting nontrivially on $x,y$
the one which becomes part of the superconformal algebra. The same
argument then shows that in this sigma model $x,y$ must not show up:
even if semiclassically there is a throat at $X=x=y=0$ through which
the three branches join, quantum mechanically they are split and there
is no way to transition from one to the other. This is exactly what
happened, in a simpler setup, in the heterotic case
\cite{Garcia-Etxebarria:2014txa}.

\medskip

To sum up, the theory perturbed by \eq{ABS} flows in the IR to a
superconformal $(4,4)$ model, with the same topological charge as
$\bC^2/\bZ_2$, and with singular behavior at the origin. It seems
fairly natural to identify this theory as type II string theory on
$\bC^2/\bZ_2$.

\section{Codimension 8 solitons are F1 strings}
\label{sec:codim-eight}

Let us consider the tachyon backgrounds associated to $\Pi_7(SO)=\IZ$, which describe  supercritical bundles with non-trivial second Pontryagin class. Using e.g. the profile (\ref{abs}) as prototype, the tachyon condensation removes the supercritical dimensions everywhere except at the origin in $\IR^8$. We will argue that, in type IIA/0A theory, the left-over real codimension-8 topological defect actually corresponds to a fundamental string (upon T-duality on a circle, the winding string turns into a momentum mode in type IIB/0B theory). In order to show it, we derive a supercritical analog of the familiar 10d IIA  1-loop term $B_2X_8$ \cite{Vafa:1995fj}, which shows that a configuration with nonvanishing second Pontryagin class in the supercritical dimensions carries the charge of an F1 string. The strategy to compute the 1-loop Chern-Simons term is to relate the computation, via a T-duality on a circle, with an anomaly cancellation computation in the supercritical type IIB theory (i.e. the $\IZ_2$ orbifold of supercritical 0B).

\subsection{The one-loop Chern-Simons term in 10d type IIA}
\label{sec:oneloopiia}

Recall that 10d type IIA theory has the one-loop Chern-Simons term for its NSNS 2-form $B_2$ \cite{Vafa:1995fj}
\begin{equation}
B_2\, X_8(R)
\label{csiia}
\end{equation}
where $X_8$ is a polynomial in the curvature 2-form $R$. It can actually be shown to correspond to the anomaly polynomial of the 6d worldvolume field theory on a type IIA NS5-brane, since the above coupling cancels this anomaly by inflow \cite{Witten:1995em}, as we now review for completeness. The coupling induces an anomalous Bianchi identity for the field strength $H_7$ of the dual gauge potential
\begin{equation}
dH_7=X_8\quad ,\quad H_7=dB_6+Q_7
\end{equation}
where $Q_7$ is the Chern-Simons form obtained from descent $X_8=dQ_7$.

The kinetic term of $B_2$, recast using the dual potential $B_6$, and field strength $H_7$, displays an anomalous variation 
\begin{equation}
S_{\rm kin}=\int_{10d} H_3\wedge  H_7\, \quad, \quad \delta S_{\rm kin}=\int_{10d} H_3\wedge  \delta Q_7= \int_{10d} H_3 \wedge dX_6 =-\int_{10d} dH_3\wedge X_6\nonumber
\end{equation}
where we have used the descent $\delta Q_7=X_6$, and integration by parts. In the presence of an NS5-brane, $dH_3=\delta_4(\rm NS)$, where $\delta_4$ is a bump 4-form transverse to the NS5-brane volume. Hence the variation is
\begin{equation}
 \delta S_{\rm kin}=-\int_{10d} \delta_4(\rm NS)\wedge X_6=-\int_{\rm NS} X_6
\end{equation}
and cancels the chiral anomaly of the NS5-brane worldvolume field theory if $X_8$ is the anomaly polynomial, as anticipated.

A simple way to compute the coupling (\ref{csiia}) is to compactify on $\IS^1$. In the resulting 9d theory, there is a coupling $A_1\, X_8(R)$, where $A_1$ arises from $B_2$ with one leg on $\IS_1$. In the T-dual IIB, the $A_1$ field is a graviphoton and the corresponding  coupling involves the KK gauge field coming from the metric. This coupling appears not from dimensional reduction of a one-loop 10d coupling, but rather as a genuinely 9d Chern-Simons coupling appearing from the parity anomaly \cite{Forte:1986em,nakahara2003geometry}. It can be easily obtained from the chiral type IIB field content, by integrating out the infinite towers of KK modes (which are charged under the KK gauge field) \cite{Liu:2013dna}. The type IIA polynomial $X_8$ is thus also given by the 9d type IIB parity anomaly, which in turn is directly related to the 10d type IIB anomaly polynomial by the descent relations.

In the coming sections, we use this logic (in reverse) to show that the supercritical IIA  has a coupling that generalizes (\ref{csiia}) and that includes the curvature of the normal bundle to the critical slice. We start with the computation of the supercritical IIB anomaly polynomial, compactify on $\IS^1$ and compute the parity anomaly by the descent relations. Upon T-duality and decompactification, this allows to obtain the Chern-Simons coupling in supercritical IIA theory. We further show that the generalized polynomial is given by the anomaly polynomial of the NS-brane of supercritical IIA, as required by (a straightforward generalization of) the above worldvolume anomaly cancellation argument.

\subsection{Anomaly cancellation in supercritical 0B}
\label{sec:anomaly-superiib}

Our starting point is the computation of anomaly cancellation of
$(10+2k)$-dimensional supercritical IIB theory (i.e. the $\IZ_2$
orbifold of the supercritical 0B theory, described in section
\ref{supercritical-intro}). The relevant sector is localized in the
$(10+k)$-dimensional fixed locus $\Sigma$ of the $\IZ_2$ orbifold;
although the field content is non-chiral with respect to $(10+k)$
diffeomorphisms, it is chiral if we also consider diffeomorphisms in the directions of the $k$-dimensional normal bundle $N\Sigma$, and there is an associated anomaly polynomial. We denote by $R$ and $F$ the curvature 2-forms of the tangent bundle $T\Sigma$ and the normal bundle $N\Sigma$. Spinors of positive chirality will be denoted by $\ket{s}$, and those of negative chirality by $\ket{c}$. 

The chiral spectrum (in the above sense) in the $(10+k)$-dimensional fixed locus of the supercritical IIB theory is:
\begin{itemize}
\item (a) Two negative-chirality spinors of $SO(10+2k)$, , with a Majorana condition, and charged in the vector representation $SO(10+k)$. The spinor decomposes as $\ket{s}\ket{c}+\ket{c}\ket{s}$ in terms of $SO(10+k)\times SO(k)$ reps. There is also a reality condition, giving a factor of 1/2 in the anomaly computation (which cancels with the fact that we have two such fields).
\item (b) Two spinors of $SO(10+2k)$, charged as a vector of $SO(k)$. These become $\ket{s}\ket{s}+\ket{c}\ket{c}$ in terms of $SO(10+k)\times SO(k)$ reps. 
\item (c) A RR self-dual $(4+k)$-form, and an anti-self dual RR $(4+k)$ form.
\end{itemize}
To compute the anomaly polynomial, we introduce the Chern characters in the vector and spinor representations of $SO(n)$, $c(F)$,$c_s(F)$, $c_c(F)$. The contributions from the different fields are
\begin{itemize}
\item (a) There are $(10+k)$-dimensional gravitinos transforming as spinors of the normal bundle, which give:
 \begin{align}-\Big[\,\hat{A}(R)\, \big[ {\rm ch}(R)-2\big] \, (c_c(F)-c_s(F))\,\Big]_{12+k}=\,\Big[\,\frac{\hat{A}(R)}{\hat{A}(F)}\,[{\rm ch}(R)-2]\, \Big]_{12}\,e(F).\end{align}
Here, we have used the identity $(c_c(F)-c_s(F))=e(F)/\hat{A}(F)$ for the Chern character for the spin cover of a bundle, with $e(F)$ being the Euler class of the normal bundle.

\item (b) There are $(10+k)$-dimensional spinors, transforming as vector-spinors of the normal bundle, and give:
\begin{align}\Big[\, \hat{A}(R)(c_c(F)-c_s(F))c(F)\,\Big]_{12+k}=-\Big[\,\frac{\hat{A}(R)}{\hat{A}(F)}\,c(F)\Big]_{12}\,e(F).\end{align}
\item Self-dual forms: These are trickier to compute, as we need to know how the orbifold affects the forms to compute the anomalies. Recalling the earlier behaviour of spinor contributions, the natural result is 

 \begin{align}\Big[\,\frac18\frac{L(R)}{L(F)}\,\Big]_{12}\, e(F).\label{anself}\end{align}
 Note that for $k=0$ this reproduces the standard result for 10d type IIB theory (using $e(F)=1$). The above generalization to arbitrary $k$ is also consistent with all our other computations. Indeed, this is the only choice ensuring cancellation of all anomalies, which are anyway located on the critical slice thanks to the $e(F)$ factor. 
\end{itemize}

The total anomaly is 
\begin{align}\left[\frac{\hat{A}(R)}{\hat{A}(F)}[ch(R)-2-c(F)]-\frac18\frac{L(R)}{L(F)}\right]_{12} \, e(F)\end{align}

In any situation leading to 10d IIB upon tachyon condensation, the bundles are such that $e(F)$ is merely localized onto the critical 10d slice. Hence, the above is the IIB anomaly polynomial, but with the characteristic classes evaluated for the $K$-theory element $(T\Sigma,N\Sigma)$ (rather than just the tangent bundle $T\Sigma$). Notice that this invariance, which we have argued for throughout the paper, depends crucially on the ansatz \eq{anself}, which also ensures anomaly cancellation and reduction of the anomaly polynomial to the 10d one when the supercritical bundles are trivial.

\subsection{Chern-Simons couplings}
\label{sec:cssuperiia}

The strategy now is to compactify supercritical IIB theory on an $\IS^1$, i.e. take supercritical 0B theory in $\IS^1\times\IR^{9+k}\times\IR^k/\IZ_2$, and to compute the parity anomaly in the lower-dimensional theory. The only contributing fields are the KK modes of those fields entering into the anomaly of the uncompactified theory, by descent from the anomaly polynomial in $10+2k$ dimensions, as explained below.

All the fields contributing to the anomaly are sections of $T\Sigma\otimes N\Sigma$. Upon KK reduction we have the splitting $T\Sigma=T\Sigma'\oplus U(1)$, while $N\Sigma$ stays the same. We will refer to the curvature and gauge potential of the $U(1)$ factor as $G$, $A_G$, and denote by $R'$ the curvature of $T\Sigma'$. 

Associated to each chiral field there is an anomaly polyform $P(R,F)$. Its degree-$(12+k)$ part is the anomaly polynomial, whereas the degree-$(10+k)$ part (modulo some factors which depend on the reality conditions for each field and which will be discussed later on) is the exterior derivative of the parity anomaly \cite{Forte:1986em,nakahara2003geometry}. The way to compute the CS term we are interested in is to expand ${\rm ch}(G)[P(R',F)]_{10+k}$, take the term linear in $G$, that is $G P(R',F)$, and then the anomaly term is $A_G P(R',F)$ where $dA_G=G$. 
We now compute these contributions from the different (KK towers of) fields:

\begin{itemize}
\item (a) For the vector-spinors of $SO(1,9+k)$ with negative chirality,  we get a term from
\begin{equation} 
P_{10+n}=e(F)\frac{\hat{A}(R')}{\hat{A}(F)}[c(R')_{9+k}-1]=\frac{\hat{A}(R')}{\hat{A}(F)}[c(R)_{10+k}-2].
\end{equation}
  We only remove the contribution of one spinor instead of two because one of the constraints (invariance under $\psi\rightarrow \psi+p\xi$, with $\xi$ a spinor field, and $p$ the momentum) is absent for a massive spinor. The KK modes are not independent due to the reality condition in 10d, so we should count only half of the KK modes (say the positive ones), but this factor will cancel with the fact that there are two identical vector-spinors. Using a zeta function regularization $\sum_{n=0}^\infty n=-1/12$, we are left with the contribution
\begin{align}
P_{8+n}^{(a)}=
\frac{1}{12}\Big[\, e(F)\,\frac{\hat{A}(R')}{\hat{A}(F)}\, \big[c(R')_{10+k}-2 \big]\,\Big]_{8+k}.
\end{align}
\item (b) Similarly, for the vector-spinors of $SO(k)$, we get an anomaly
\begin{align}
P_{8+k}^{(b)}=-\frac{1}{12}\Big[\, e(F)\, \frac{\hat{A}(R')}{\hat{A}(F)}\,c(F) \Big]_{8+k}.
\end{align}
\item (c) The self-dual forms are trickier, but the natural guess is that they mimic an anti-self dual form for which the Chern character factor coming from the $N\Sigma$ contribution is precisely $\frac{e(F)}{L(F)}[\Sigma']c(G)$. With this proviso, we may analyze the parity anomaly. For the $(4+2k)$-form, we get a tower of complex massive KK forms in $9+2k$ dimensions. The reality condition restricts the sum to positive momenta, and we also drop the factor of $1/2$ in the anomaly polynomial since the 9d $(4+2k)$-forms are complex. Finally, the dimension of the Dirac spinor in $9+2k$ dimensions is the same as that of one Weyl spinor in $10+2k$, so unlike in standard anomaly cancellation we don't have to divide the index by two. So the parity anomaly is obtained through descent procedure from the chiral anomaly of a complex anti self-dual $(4+2k)$-form in $10+2k$ dimensions, namely $-\frac12 L(R)$. Adding up all the states in the tower plus the normal bundle contribution we get a contribution
\begin{align}
P_{8+n}^{(c)}=-\frac{1}{24}\Big[\,\frac{L(R')}{L(F)}\,e(F)\Big]_{8+n}.\end{align}
\end{itemize}
The total parity anomaly is
\begin{align}
P_{8+k}=P_{8+k}^{(a)}+P_{8+k}^{(b)}+P_{8+k}^{(c)}=\frac{e(F)}{12}\left[\frac{\hat{A}(R')}{\hat{A}(F)}\, \big[\, c(R)_{10+k}-2-c(F)\,\big]-\frac12\frac{L(R')}{L(F)}\right]_{8}.
\end{align}

Notice this nicely  respects the $K$-theory invariance of the class $(T\Sigma',N\Sigma)$, as it only involves characteristic classes of the difference bundle.

The parity anomaly induces a Chern-Simons coupling $A_G P_{8+n}(R',F)$ in the supercritical type IIB theory. Upon T-duality on the $\IS^1$, the gauge field $A_G$ becomes a component of the higher-dimensional NSNS 2-form $B_2$, so the above term arises from the KK reduction of a Chern-Simons coupling in  the supercritical IIA theory of the form
\begin{equation}
B_2 \,  P_{8+n}(R,F)
\label{superc-cs}
\end{equation}
Note that for $F$ trivial we recover the familiar $B_2X_8$ term of 10d type IIA theory (\ref{csiia})

\subsection{NS-brane worldvolume anomaly cancellation}
\label{sec:nsbrane-anomaly}

The 10d type IIA CS coupling $B_2 X_8$ is a crucial ingredient in the cancellation of the NS5-brane worldvolume 6d anomalies by inflow, and explains that $X_8$ is actually the anomaly polynomial of this 6d theory \cite{Witten:1995em}. In a similar spirit, we will show that the above CS coupling in the supercritical IIA theory  precisely produces the inflow cancelling the worldvolume anomalies of the NS$(5+2k)$-brane (the object charged magnetically under $B_2$).

The chiral field content on the worldvolume of an NS$(5+2k)$-brane in
supercritical type IIA theory can be guessed by the requirement that
upon bulk closed tachyon condensation it becomes the standard
NS5-brane worldvolume field content of 10d type IIA theory. Namely,
this spectrum consists of two fermions of same chirality in 6
dimensions, plus a self-dual 2 form. The fermions should arise out of
localized modes of supercritical fermions in the tachyon background,
as briefly discussed in section \ref{kk-codim4} and argued in
\cite{Hellerman:2004zm}. As for the self-dual form, in the critical
theory it arises as the reduction of the bulk anti self-dual 4-form
along the anti self dual 2-form of the Taub-NUT background. One may
expect the same logic to hold in the supercritical picture, so one
expects to have a self-dual $(4+k)$-form in the supercritical theory,
along with other lower $p$-forms. This sector will, upon tachyon
condensation, yield the self-dual 2-form and one scalar.

 In what follows we prefer to provide a more direct and rigorous derivation, from T-duality with a Taub-NUT configuration in supercritical IIB (which is easily understood to hold as in the critical case e.g. using Buscher's worldsheet derivation of T-duality).
In fact, the set of localized modes at the center of Taub-NUT spaces can be obtained, for $k$ coincident centers, by replacing the near-center region of the Taub-NUT by an $A_{k-1}$ singularity, i.e. a $\IC^2/\IZ_k$ orbifold. Since we are interested in the chiral content, the anomaly is given by $n$ copies of that in a single Taub-NUT center (i.e. a single NS-brane in the type IIA theory). 

For simplicity, we focus on the supercritical type IIB theory with a $\IC^2/\IZ_2$ along the critical slice. 
Since supercritical IIB is already a $\IZ_2$ orbifold of supercritical 0B, we have to compute the spectrum in a $\IZ_2\times\IZ_2$ orbifold.  As in section \ref{supercritical-intro} we denote by $g$ the generator of the $\IZ_2$ to become the IIB GSO projection; we also use $h$ for the additional $\IZ_2$ action $X\rightarrow -X$ on four of the 10d coordinates, and their worldsheet superpartner fermions. Let us now discuss the spectrum in sectors twisted by $h$ and $gh$, which lead to the localized matter:
 \begin{itemize}
\item $h$-twisted sector:  This contains NSNS and RR sectors, and chiral matter arises only from the latter. The zero point energies vanish, and there are fermion zero modes in all directions except those twisted by $h$. Massless states are the tensor product of two $SO(6+2k)$ spinors (with same relative chirality in the 0B theory). This produces a set of even-degree RR $p$ form potentials, including a chiral $(2+k)$-form, self-dual in the $(6+2k)$-dimensional fixed locus. These forms can be regarded as arising from the bulk RR forms supported on the harmonic 2-form associated to the collapsed 2-cycle, e.g. the self-dual $(2+k)$-form arises from integrating the bulk self-dual $(4+k)$-form over the 2-cycle.

Using the latter interpretation it is clear that the contribution to the anomaly polynomial is
\begin{align}
X^{(d)}_{8+k}=-\frac18\left[\frac{L(R)}{L(F)}\,e(F)\right]_{8+k}.
\end{align}

\item $gh$-twisted sector: This contains NS-R and R-NS sectors, and behaves as a $g$-twisted sector with $k+4$ twisted dimensions instead of just $k$. The zero point energies vanish in both the NS and R sectors, so massless states arise from degenerate groundstates from the fermion zero modes.

The spectrum consists of two $(6+2k)$-dimensional spinors of $SO(1,2k+5)$ of definite chirality, transforming as $(1/2,0)$ spinors of the $SO(4)$ normal to the fixed locus. The anomaly polynomial of the chiral matter of this sector is
\begin{align}X^{(e)}_{8+k}=2\left[\frac{\hat{A}(R)}{\hat{A}(F)}e(F)\right]_{8+k}.\end{align}
\end{itemize}
The total worldvolume anomaly polynomial of the $NS(5+2k)$-brane is
\begin{align}
X^{NS}_{8+k}=X^{(d)}_{8+k}+X^{(e)}_{8+k}=e(F)\left[2\frac{\hat{A}(R)}{\hat{A}(F)}-\frac18\frac{L(R)}{L(F)}\right]_{8}
\end{align}
We have the amazing fact that $X_{8+n}^{\rm NS}=P_{8+n}$, ensuring cancellation of anomalies by inflow from the Chern-Simons coupling (\ref{superc-cs}). This is just the ordinary cancellation condition for the 8-form part of IIB anomaly polynomial, but the characteristic classes evaluated at the K-theory classes.

\subsection{Other closed-string tachyon solitons}
\label{sec:other}

In the previous subsections, we have described the endpoint of tachyon condensation for profiles carrying nontrivial charges in the integer-valued real K-theory groups $KO(\IS^4)$ and $KO(\IS^8)$. We have also provided evidence of the K-theory equivalence picture between the supercritical bundles upon tachyon condensation, for instance verifying that the anomaly polynomial of supercritical IIB  or the $B_2X_{8+2k}$ coupling in supercritical IIA depend on the supercritical dimension bundles only through their K-theory class.

Keeping up with this picture, one should also be able to study solitons charged under the $\IZ_2$-valued $K$-theory groups $KO(\IS^k)$ for $k=1,2,9,10$. Such objects are necessarily non-supersymmetric, and therefore much less understood than their integer-valued counterparts which, as we have seen, correspond to familiar objects in type II string theory. The great advantage of the K-theory description of the supercritical theory is that, once established, it allows us to describe these stable but exotic solitons on the same footing as the more well-known ones. 

\begin{itemize}
\item Associated to $KO(\IS^{10})$ we have a $\IZ_2$ instanton. In the type II theory, this can be constructed as a gravitational instanton associated to the first factor of the homotopy group $\pi_9(SO)=\IZ_2\oplus\IZ$ of the tangent bundle. The second factor is unstable, meaning that a representative of a nontrivial class of it can be made trivial by the addition of extra supercritical dimensions. 

To our knowledge, this non-supersymmetric object has not been described before in the literature. It is a gravitational instanton which in principle should provide nonperturbative contributions to the 10-dimensional type II couplings, albeit to higher-derivative terms due to many fermionic zero modes due to the broken supersymmetries.

\item For $KO(\IS^9)=\IZ_2$ we have a non-BPS particle in both type II theories. It is clearly different from the familiar ($\IZ$-valued) type IIA D0-brane, or the (uncharged non-BPS) type IIB $\hat{D0}$-brane. Standard analysis as in \cite{Witten:1998cd,Gukov:1999yn} suggest it transforms as a spinor (of both the supercritical bundles, and of the critical spacetime). As we argue later, the $\IZ_2$ charge is fermion number mod 2, so it is possible that this particle simply decays into a perturbative spinor. If not, it would be interesting to follow this particle in dual pictures, in particular in the M-theory lift of type IIA.

\item For $KO(\IS^2)$, we get the dual string for the particle described above. Before the tachyon condensation it corresponds to a geometric configuration in which an $\IR^2$ in the supercritical dimensions picks up a $2\pi$ rotation as one moves around the  core of the soliton. This a non-trivial action on the spacetime fermions, which survives even after tachyon condensation. Therefore the string is associated to a $\IZ_2$ discrete (gauge) symmetry under which only fermions are charged. It also acts non-trivially on the non-BPS particle described above, in nice agreement with our statement that it is a spinor (and thus a fermion).

\item Finally, $KO(\IS^1)$ corresponds to a domain wall. Before the condensation, it corresponds to a geometric configuration with one supercritical dimension fibered as in a Moebius strip. In analogy with the heterotic setting \cite{Garcia-Etxebarria:2014txa} it seems reasonable to conjecture that amplitudes containing an instanton and a domain wall will get an extra phase when crossing each other.

\end{itemize}

We leave further discussion of these objects for future work.

\section{A supercritical viewpoint on GLSMs}
\label{sec:glsm}

Non-trivial Calabi-Yau threefolds are often constructed as (complete
intersections of) hypersurfaces in toric varieties. When these
Calabi-Yau threefolds are taken to be backgrounds for type II string
theory, the toric construction admits a beautiful realization in
string theory \cite{Witten:1993yc}. Instead of trying to construct the
CFT describing the non-linear type II sigma model in the target
Calabi-Yau, we construct a particular non-conformal field theory in
the worldsheet, the $(2,2)$ \emph{Gauged Linear Sigma Model}
(GLSM). At low energies this field theory will flow to the non-linear
sigma model, so we recover the previous physics, but the UV
description makes the geometric construction as a hypersurface
manifest. (Remarkably, the GLSM description does, in addition, allow
for exploration of the aspects of string theory not captured by
classical geometry, but this will not concern us here.)

A simple example is the quintic Calabi-Yau threefold, which can be
constructed as a degree five hypersurface in the ambient toric space
$\bP^4$. In the GLSM description a slightly different description
arises. We consider a $U(1)$ gauge group in our $(2,2)$ worldsheet
theory, and take a set of six (complex) chiral superfields $z_i,P$
with charges
\begin{equation}
  \label{eq:quintic-GLSM}
  \begin{array}{c|cccccc}
    & z_1 & z_2 & z_3 & z_4 & z_5 & P\\
    \hline
    U(1) & 1 & 1 & 1 & 1 & 1 & -5
  \end{array}
\end{equation}
In order to be in the geometric phase of the quintic, we also choose a
value for the Fayet-Iliopoulos term for the $U(1)$ which forces
$\sum |z_i|^2>0$, or in other words it disallows all the $z_i$ from
vanishing simultaneously. We introduce, in addition, a superpotential
of the form
\begin{equation}
  \label{eq:quintic-W}
  W_{GLSM} = f_5(z_i) \, P
\end{equation}
where $f_5(z_i)$ is a homogeneous polynomial of degree 5 in the
$z_i$. We now look to the classical set of vacua for this theory. The
set of fields in~\eqref{eq:quintic-GLSM}, plus the D-term constraint,
parameterize a Calabi-Yau fivefold given by the $\cO(-5)$ bundle over
$\bP^4$. The F-terms coming from~\eqref{eq:quintic-W} then impose
$f_5(z_i)=0$, and for smooth quintics (i.e. such that $df_5=f_5=0$ has no
solutions) also $P=0$. Geometrically, it is most natural to take $P=0$
first, which restricts us to the $\bP^4$ base of the fibration, and
then $f_5(z_i)=0$ reproduces the classical description of the quintic.

This construction will be familiar to most readers, but given the
focus of this paper we would like to reformulate it from the following
supercritical type II viewpoint. We consider the GLSM
in~\eqref{eq:quintic-GLSM}, initially with no superpotential, as
describing a supercritical type 0 on the resulting $\cO(-5)\to\bP^4$
fivefold (therefore suitably dressed with a linear dilaton
background). To connect with type II, we impose a $\bZ_2$ quotient
chosen to act as $P\to -P$, leaving the $z_i$ invariant. Notice that
differently to previous cases, we make no distinction between
``critical'' and ``supercritical'' $z_i$ coordinates, all of them
enter equally in the construction.\footnote{Up to the same point in
  the argument, morally the same was true in our previous cases, since
  the distinction between $x$ and $X$ coordinates only mattered once
  we chose a tachyon profile and a branch of the resulting moduli
  space. In the current case, due to the non-trivial topology, it
  makes no sense to partition the $z_i$ coordinates, even in
  principle.}

Type 0 on the fivefold has a real closed string tachyon. We choose a
profile for this tachyon given by
\begin{equation}
  \label{eq:quintic-tachyon}
  \tach = \re(W_{GLSM}) = \re(f_5(z_i) \, P)\, ,
\end{equation}
ignoring as usual the lightlike profile. The solution
of the equations of motion at late times is the subvariety cut out by
\begin{equation}
  \label{eq:supercritical-GLSM-eom}
  \frac{\partial\tach}{\partial
    x_k}=\frac{\partial\tach}{\partial
    y_k}=\frac{\partial\tach}{\partial
    p}=\frac{\partial\tach}{\partial q} = 0
\end{equation}
where we have introduced $z_k=x_k+iy_k$, $P=p+iq$. (Some subtler
details of the dynamics of this system were worked out in
\cite{Hellerman:2006ff}.) Since $W_{GLSM}$ is holomorphic in $z_i,P$,
one easily sees that~\eqref{eq:supercritical-GLSM-eom} is equivalent
to\footnote{Here is an argument: write $W_{GLSM}=u+iv$. Since it is
  holomorphic on $z_i,P$, one has the Cauchy-Riemann conditions
  $\frac{\partial u}{\partial x_k}=\frac{\partial v}{\partial y_k}$,
  $\frac{\partial u}{\partial y_k}=-\frac{\partial v}{\partial x_k}$,
  and similarly for $P$. The supercritical equations of motion
  \eqref{eq:supercritical-GLSM-eom} are just
  $\frac{\partial u}{\partial x_k} = \frac{\partial u}{\partial y_k} =
  \frac{\partial u}{\partial p} = \frac{\partial u}{\partial q} = 0$.
  These equations together imply~\eqref{eq:GLSM-eom}, if we expand it
  in real components. The reverse implication is also straightforward:
  using the real decomposition of
  $\frac{\partial W_{GLSM}}{\partial z_k}=0$, together with
  holomorphy, one gets
  $\frac{\partial u}{\partial x_k} = \frac{\partial u}{\partial y_k} =
  0$.}
\begin{equation}
  \label{eq:GLSM-eom}
  \frac{\partial W_{GLSM}}{\partial z_k} = \frac{\partial
    W_{GLSM}}{\partial P} = 0\, .
\end{equation}
In words, we perfectly reproduce the GLSM construction, and end up at late
times with ten-dimensional type II string theory on
$P=f_5(z_i)=0$ (for smooth $f$; see later for comments on singular cases).

Besides providing an interesting alternative viewpoint on the usual
GLSM construction, the supercritical viewpoint has an interesting
application to the study of discrete symmetries in string theory. The
idea is a straightforward generalization of the discussion in
\cite{Berasaluce-Gonzalez:2013sna}: before the tachyon starts
condensing, we have a rather large continuous symmetry group acting on
$\cO(-5)\to\bP^4$ (in particular acting on the $\bP^4$). The choice of
the tachyon~\eqref{eq:quintic-tachyon} breaks down this continuous
group into a discrete subgroup for appropriate choices of
$f_5(z_i)$. For instance, take the Fermat quintic
\begin{equation}
  f_5(z_i) = z_1^5 + z_2^5 + z_3^5 + z_4^5 + z_5^5 -\psi z_1z_2z_3z_4z_5
\end{equation}
where $\psi$ is a complex structure parameter, which we take to be
nonvanishing. In this case we have a non-abelian symmetry group given
by $(S_5\ltimes(\bZ_5)^4)/\bZ_5$ with $S_5$ the permutation group of 5
elements (see for example \cite{Doran:2007jw} for a detailed
discussion of this point, and generalizations). This groups survives
in the final type II background, and has now a natural embedding into
the isometry group of the original non-compact Calabi-Yau fivefold.

The embedding of the discrete symmetries into a continuous group
allows for a direct construction of the 4d charged strings (see
\cite{Berasaluce-Gonzalez:2013sna} for a related but different
approach). For a given $\IZ_p$ string, one simply constructs a vortex
of the Kibble mode associated to the corresponding $U(1)$ symmetry in
the continuous group. It would be interesting to explore the string
creation effects associated to mutually non-commuting strings.

It is also interesting to consider what happens to the background when
the discrete symmetry gets enhanced to a continuous one, i.e. when we
approach the core of the vortex. Choose for instance a continuous
deformation of the Fermat quintic taking us to the singular space
$X_s$ given by $f_5(z_i)=\psi z_1z_2z_3z_4z_5$. In this limit the
discrete symmetry is enhanced to a continuous group containing
$(S_5\ltimes U(1)^5)/U(1)$. Topologically $X_s$ is a set of 5
intersecting $\bP^3$s given by $z_i=0$. This manifold is singular, and in fact it is an example of the ground state varieties with non-transversal polynomials described in \cite{2005JGP....53...31H}. Classically, in addition to the non-linear sigma-model branch, the
$P\neq 0$ branch of the CFT opens up at the singularities,
i.e. whenever two $\bP^3$ components intersect. In related contexts
\cite{Witten:1994tz,Strominger:1995cz} there is evidence that the
appearance of these branch structure signals the breakdown of
worldsheet CFT, and that we need to include the effects of light brane
states. This is also the case here. In the original quintic we had 204
independent homology three-cycles. These all become localized at the
singularities of $X_s$, since $\bP^3$ does not support any non-trivial
three-cycle. In the type II theory the branes wrapping these cycles
will become massless. Notice that differently from the usual conifold
transition, here both electric cycles and their magnetic duals are
becoming massless, so we expect string theory in this background to be
described in four dimensions by a theory without a Lagrangian
description.

\medskip

The reinterpretation of the GLSM generalizes easily to spaces beyond the quintic. As
a simple example, let us consider the non-compact Calabi-Yau sixfold
given by the GLSM
\begin{equation}
  \begin{array}{c|cccccccc}
    & z_1 & z_2 & z_3 & z_4 & x & y & z & P\\
    \hline
    U(1)_1 & 1 & 1 & 1 & 1 & 8 & 12 & 0 & -24\\
    U(1)_2 & 0 & 0 & 0 & 0 & 2 & 3 & 1 & -6
  \end{array}
\end{equation}
Choose as above $\tach=\re(W_{GLSM})$ with
\begin{equation}
  W_{GLSM} = P\bigl(y^2 + x^3 + f(z_i) xz^4 + g(z_i)z^6 \bigr)
\end{equation}
which after tachyon condensation, for a suitable choice of FI terms,
leaves us with an elliptically fibered fourfold, with base
$\bP^3$. For homogeneity we need that $f(z_i)$ and $g(z_i)$ are
homogeneous polynomials of $z_i$ of degrees 16 and 24 respectively. If
we want to engineer a background with non-abelian discrete symmetries,
we could choose for instance, in analogy with our discussion for the
quintic
\begin{align}
  f(z_i) & = \sum_{i=1}^4 z_i^{16} + \psi_1 (z_1z_2z_3z_4)^4\\
  g(z_i) & = \sum_{i=1}^4 z_i^{24} + \psi_2 (z_1z_2z_3z_4)^6
\end{align}
which are invariant under a manifest symmetry group
$S_4\ltimes \bZ_2^4$ acting on the $\bP^3$ coordinates, leaving
$(x,y,z,P)$ invariant. Richer configurations can appear if we can
induce discrete symmetries acting non-trivially on the fiber, which
becomes particularly easy if we choose different realizations of the
torus fiber \cite{Klemm:1996hh,Aldazabal:1996du}, such as a cubic in $\bP^2$, or a quartic on
$\bP^{112}$. This construction has potential applications to the
realization of non-abelian symmetries in F-theory.\footnote{See
  \cite{Braun:2014oya,Morrison:2014era,Anderson:2014yva,Klevers:2014bqa,Garcia-Etxebarria:2014qua,Mayrhofer:2014haa,Mayrhofer:2014laa,Cvetic:2015moa}
  for recent work on abelian discrete symmetries in F-theory, and
  \cite{Grimm:2015ona} for work on non-abelian discrete symmetries in
  F-theory.}

\medskip

Another interesting observation in this context is that closed string
tachyon condensation on flat space can give an alternative
supercritical construction of ADE singularities for appropriate
choices of the tachyon profile. For instance, start with $\bC^4$
parameterized by $(z_1,z_2,z_3,P)$, with the $\bZ_2$ orbifold acting
as $(z_i,P)\to(z_i,-P)$. We choose a tachyon profile
$\tach=\re\bigl((z_1^2+z_2^2+z_3^2)P\bigr)$. After the tachyon
condenses we end up with $z_1^2+z_2^2+z_3^2=0\subset \bC^3$, a well
known description of $\bC^2/\bZ_2$.

\section{Supercritical bundles for F-theory matrix factorizations}
\label{sec:fth}

In \cite{Collinucci:2014taa,Collinucci:2014qfa}, type IIB tachyons describing particular backgrounds were related to particular  \emph{matrix factorizations} (MF) of polynomials living on an ambient auxiliary space sharing some resemblance to the total space of F-theory fibrations. The matrices of the matrix factorization are sections of bundles defined over the ambient space, much like our tachyon matrix $\itach$ above is a section of the $x,y$ bundles over the critical slice.  It is therefore tempting to try to give a physical meaning to this ambient space, and the associated matrix factorizations, by somehow reinterpreting it as a supercritical type 0 background, with tachyon condensation reducing again the dynamics to the type 0 critical slice (the ambient space). 

We will first briefly review the recipe of \cite{Collinucci:2014taa}
in the type F-theory context. An F-theory geometry is commonly
specified by a possibly singular Calabi-Yau fourfold $\IX_4$ with a
torus fibration. This fourfold is part of an M-theory background
$\IX_4\times \IR^{2,1}$ which is related to the four-dimensional
F-theory model by taking a limit in which the size of the torus fiber
shrinks to 0. Enhanced gauge groups and chiral matter may arise at
singularities of the fiber.
As discussed above, a convenient class of backgrounds $\IX_4$ can be constructed in terms of
a hypersurface $f=0$ on an ambient toric space $\mathcal{A}$.

The polynomial $f$ specifies the geometry of the fibration. However, this does not fully specify the M/F-theory background; the profile of the M-theory 3-form $C_3$ must also be specified. For backgrounds without $G$-flux the choice of $C_3$ is equivalent to picking a particular element in the Deligne cohomology of $H_3(X_4,\IZ)$ \cite{Donagi:2011jy}. For singular fourfolds, there may be vanishing cycles on which one can wrap $C_3$; these backgrounds typically correspond to T-branes in the type IIB setup \cite{Donagi:2011jy}.

The proposal of \cite{Collinucci:2014taa} is that these ``T-brany''
degrees of freedom of the singular fourfold are captured by the
different matrix factorizations of the polynomial $P$. A $n\times n$
MF of a polynomial $P$ is a pair $(A,B)$ of matrices satisfying
\begin{equation}
  \label{eq:factorization}
  A\cdot B=B\cdot A = f \cdot \mathbf{I}_{n\times n}.
\end{equation}
A particular (stable) matrix factorization gives a particular T-brane
background. The reverse is not quite true: a T-brane background modulo
complexified gauge transformations corresponds to a particular
\emph{equivalence class} of matrix factorizations (which furnish the
so-called category of \emph{stable} matrix factorizations).

The MF is naturally associated to a complex of vector bundles over the ambient space $\mathcal{A}$, which is formally very similar to those appearing in open tachyon condensation in brane-antibrane systems. In the present setup, however, there are no open string sectors. This seems to demand a setup in which there is there are bundles associated to geometry, and a mechanism to annihilate them. It is very tantalizing to suggest that the tachyon condensation of bundles in the complex defined by the MF is physically realized in terms of supercritical strings. In the following we give some suggestive hints in this direction.

A first step is to connect F-theory to the supercritical string setups considered in the paper. This can be done by defining F-theory  from M-theory on $\IX$ as described above, and further compactifying on an extra $\IS^1$ to connect with type IIA on $\IX$.  Conversely, we start with type IIA on an elliptically fibered  variety $\IX$, lift to M-theory on $\IX\times \IS^1$, decompactify the $\IS^1$, and subsequently\footnote{We do not consider other limiting procedures, which may lead to other  possibly interesting outcomes.} shrink the elliptic fiber to connect to F-theory on $\IX$. As usual, holomorphic information is preserved under these operations, so the physics of F-theory at singularities of $\IX$ will be directly described by the physics of type IIA at singularities of $\IX$.

We can describe these type IIA compactifications in terms of a $(2,2)$ GLSM with ambient space ${\mathcal A}$ and superpotential determined by the defining polynomial $f$. In this context, there is a natural proposal to realize the inclusion of the degrees of freedom of the MF. A typical trick to introduce extra bundles is to introduce extra $(2,2)$ multiplets in the GLSM, specifically $n$ pairs of multiplets $X_a$, $Y_a$, with $a=1,\ldots, n$ (possibly charged under the 2d gauge group).  Alternatively, these can be physically understood as extra supercritical dimensions\footnote{This is in addition to possibly including the normal dimension to $\IX$ in ${\mathcal A}$ as supercritical, as in section \ref{sec:glsm}.} in a supercritical type 0 theory (or a supercritical type II theory, by simply orbifolding by $Y\to -Y$). In this language, the interpretation of the complex of bundles associated to the MF as a tachyon condensation motivates the introduction of closed tachyon background in the supercritical theory given by 
 \beqa
\tach = \re(\, \sum_{ab} X_a M_{ab} Y_b\,) 
 \label{tachyon-supo}
 \eeqa
Namely, the matrix $M$ in the MF plays as the tachyon matrix $\itach$. This tachyon background leads to the annihilation of the extra supercritical dimensions, except at special loci at which the rank of the matrix drops, which occur precisely at singular points of $\IX$. From the viewpoint of the supercritical worldsheet description, there are extra dimensions left over at these loci, which describe singular CFTs as we have described in earlier sections. A careful study
of such singular CFTs should reproduce the formal prescription of
looking to matrix factorizations~\eqref{eq:factorization}. (Not
entirely unlike how type II CFT reproduces the classification of
D-branes by matrix factorizations \cite{Herbst:2008jq}, although we
expect the details to differ significantly).

Some interesting hints from translating results of the previous sections to this setup are that complex codimension 2 defects are associated to the singular CFT of $A_n$ singularities, and therefore are naturally associated to the appearance of enhanced gauge symmetries. Also, complex codimension 4 defects are associated to the presence of F1's in the type IIA picture, which maps under the dualities to the presence of D3-branes in the F-theory setup; this nicely connects with the presence of (anti) D3-brane sources detected in \cite{Collinucci:2014taa,Collinucci:2014qfa} by the presence of matter localized at points.

Clearly, there are many points that deserve further clarification. For instance, the fact that F-theory MFs are defined by the stable category requires that widely different tachyon backgrounds lead to equivalent physics, at least in what concerns the singular structure of the model. Also, we have been deliberately ambiguous about whether the supercritical string theory is defined on $\IX$ or on $\mathcal{A}$ (times the extra `bundle' dimensions). Finally, it is not clear what computations in the (eventually singular) CFT could match the remarkably tractable computation of non-perturbative spectra in \cite{Collinucci:2014taa,Collinucci:2014qfa} in terms of exact sequences. We hope to address the
fine points of this construction in upcoming work.

\section{Conclusions}
\label{conclusions}

In this paper we have described closed tachyon solitons in the supercritical extensions of type II strings, and matched them with different defects in critical 10d type II theories. The solitons are classified by the real K-theory groups KO, for bundles associated to extra pairs of supercritical dimensions.

In contrast to the similar analysis in the heterotic context \cite{Garcia-Etxebarria:2014txa}, where supercritical solitons gave NS5-branes upon tachyon condensation, we
find evidence that in the type II context the codimension four solitons are more naturally associated with gravitational
instantons (of course, T-dual to NS5-branes in setups with one $\IS^1$ dimension). We have discussed in
particular the tachyon associated with the ABS construction, which we
have argued gives rise to a local $\bC^2/\bZ_2$ singularity upon tachyon condensation.

We have also shown that a codimension 8 soliton gives rise to a
fundamental string. Along the way we found a coupling in the
supercritical string of the form $B_2\wedge X_{8+n}$ in supercritical
IIA, which generalizes the familiar $B_2\wedge X_8$ coupling in IIA,
and which was crucial in the argument.

The K-theory viewpoint suggests that other solitons are possible, and
we have briefly discussed some of their properties. It would be quite
interesting to understand these better, and in particular to follow them under diverse dualities.

We have also given an alternative description of gauged linear sigma
models in the context of the supercritical string, in which the
ambient space becomes a physical supercritical background, and the
reduction to the Calabi-Yau is due to tachyon condensation. This
allowed us to construct representatives for the strings charged under
the discrete symmetries associated with diffeomorphisms of the
Calabi-Yau, and shed some light on the enhancement of discrete
non-abelian symmetries to continuous symmetries.

Finally, we sketched how supercritical strings may illuminate the
matrix factorization description of F-theory in singular spaces, hopefully paving the way towards a physical understanding of this prescription.

\bigskip

\section*{Acknowledgments}

We thank Mikel Berasaluce-Gonz\'alez, Andr\'{e}s Collinucci, Mirjam Cveti\v{c}, Diego Regalado, Ander Retolaza, Raffaele Savelli and Gianluca Zoccarato for useful discussions and comments. MM thanks the Max Planck Institute for Physics for hospitality during the later stages of the project. MM and AU are
partially supported by the grants FPA2012-32828 from the MINECO, the
ERC Advanced Grant SPLE under contract ERC-2012-ADG-20120216-320421
and the grant SEV-2012-0249 of the ``Centro de Excelencia Severo
Ochoa'' Programme. MM is also partially supported by the COST Action MP1210. IGE would like to thank N.~Hasegawa for kind
encouragement and support. MM is supported by a ``La Caixa'' Ph.D
scholarship.

\appendix

\section{Explicit construction of $SO$ tachyons}
\label{appA}

\subsection{Realifying the ABS construction}
\label{app:realification}
The Atiyah-Bott-Saphiro construction \eq{abs} provides a
representative of the generator of the complex K-theory group
$K(S^k)$. We are interested however in generators of the real K-theory
groups $KO(S^k)$. In this appendix we discuss how to construct one
from the other. Throughout this section $k=2p$ will be the dimension of
the transverse space to the soliton we wish to construct.

The main tool we will use is the embedding from $U(n)$ to $SO(2n)$
given by treating real and imaginary parts independently (this
embedding process sometimes goes under the name
``realification''). Given any $n$-dimensional complex vector bundle,
it can be regarded as a $2n$-dimensional real vector bundle via the
mapping
\begin{equation}
  r:\ (z_1,z_2,\ldots)\quad\rightarrow\quad
  (x_1,y_2,x_2,y_2,\ldots),\quad\text{where}\quad z_k\equiv
  x_k+iy_k.
\end{equation}
A mapping $M$ between two complex vector bundles $E,F$, such as the
tachyon, induces a mapping $M_{\IR}=r_F\circ r_E^* M$ from $r(E)$ to
$r(F)$. Given $M$, $M_{\IR}$ is just obtained by the replacement
\begin{equation}
  1\rightarrow \begin{pmatrix}1&0\\0&1\end{pmatrix},\quad
  i\rightarrow\begin{pmatrix}0&-1\\1&0\end{pmatrix}.\label{fgetcomp}
\end{equation}
in the matricial expression for $M$. If the complex bundles are $U(n)$
bundles, their realifications are $SO(2n)$ bundles: At the level of
the algebra, one can see that \eq{fgetcomp} sends a $n\times n$
anti-hermitian traceless matrix to a $2n\times 2n$ real antisymmetric
matrix, so that it indeed defines an algebra embedding from
$\mathfrak{u}(n)$ to $\mathfrak{so}(2n)$. Since $U(n)$ is a connected
Lie group, the algebra embedding extends to a group embedding via the
exponential mapping.

We may now take the tachyon \eq{ABS} (which for each point with $\vert\vec{X}\vert=1$ defines an element of $U(2^{p-1})$)  and turn the mapping between $U(2^{p-1})$-bundles to a mapping between $SO(2^p)$ bundles over the same base. However, we now have to show that the maps constructed in this way indeed represent nontrivial elements of the $KO$ groups. 

 To do this, we can now complexify again the bundles (which, for a given real bundle $E_\IR$, means taking the tensor product with the complex numbers $E_\IR\otimes\IC $). A generic section of the complexification of $r(E)$ is of the form
\begin{equation}
  s_{E_x}+is'_{E_x}+s_{E_y}+is'_{E_y}=(s_{E_x}+is'_{E_y})+i(s'_{E_x}-is_{E_y})=e+\ov{e}',
\end{equation}
where $E_x$ and $E_y$ are the real and imaginary parts of $E$, $s$
and $s'$ are sections of the corresponding bundles, and $e$ and $\ov{e}'$ are sections of $E$ and $\ov{E}$ respectively. Thus the complexification is isomorphic to $E\oplus\ov{E}$.

The complexification of $r(E)$ is trivial if $r(E)$ is or, in other
words, $r(E)$ can only be trivial if $E\oplus \ov{E}$ is.  The above discussion carries over to pairs of bundles
$(E,F)$ and to $K$ theory classes in a straightforward
manner, using the map from tachyon profiles to $K$-theory classes described in the main text. If $T$ represents the $K$-theory class $(E,F)$, then $T\oplus\ov{T}$ represents the $K$-theory class $(E\oplus\ov{E},F\oplus\ov{F})$. Therefore, if the $K$-theory element represented by
$T\oplus\ov{T}$ turns out to be nontrivial, then the realification of
$T$ describes a nontrivial element in $KO$.

In $k=4,8$ dimensions, the ABS tachyon has nontrivial $(k/2)$-th Chern class. Since $c_k(\ov{E})=(-1)^kc_k(E)$, this means that $T\oplus\ov{T}$ has nontrivial $(k/2)$-th Chern class (in fact, it is twice that of the ABS tachyon). This means that the real tachyons obtained from these indeed describe nontrivial elements of $KO$. Since the $k$-th Pontryagin class of a real bundle is defined as the $(k/2)$-th Chern class of its complexification, we see that the Pontryagin classes of these tachyons are even.

In $k=2$ dimensions $T\oplus \ov{T}$ is a trivial bundle so the above considerations do not apply, and in $k=1$ the ABS construction (which strictly speaking is only defined for even dimensions) does not apply. We will consider these cases in detail below. Notice that once we have constructed generators for the $KO$-theory groups from $k=1$ to $8$ we can get generators for any other $KO$-group via Bott periodicity \cite{Witten:1998cd}.

We now turn to the explicit construction of real tachyons using the above recipe.

For $k=4$, the ABS construction provides a tachyon, written in terms
of $\Gamma$ matrices

\begin{align}T=\Gamma^\mu X_\mu\label{tapp-Gamma}\end{align}
where $\Gamma^\mu$ are $4\times 4$ $SO(4)$ Dirac matrices. However, as stated in the text, we should regard the above tachyon as a map between chiral spinor bundles $S^\pm$, which are complex 2-dimensional. This means that a more appropriate $2\times2$ form of the above tachyon may be obtained. Define the following matrices:
\begin{align}
  \label{eq:sigma-matrices}
  \sigma^\mu & = (1,i\tau^1, i\tau^2, i\tau^3)\\
  \ov\sigma^\mu & = (\sigma^\mu)^\dagger = (1, -i\tau^1, -i\tau^2, -i\tau^3)
\end{align}
with $\tau^i$ the Pauli matrices. Then, when mapping chiral spinor
bundles, \eq{tapp-Gamma} becomes
\begin{equation}
  T=\sigma^\mu X_\mu\label{tapp1}.
\end{equation}

Applying the homomorphism \eq{fgetcomp}, we get the $4\times 4$ real
tachyon
\begin{equation}
  T=r(\sigma^\mu) X_\mu = \ov\eta_\mu\check{X}^\mu
  \label{REALabs}
\end{equation}
with $\check{X}^\mu=(X^0,X^3,-X^2,X^1)$ and $\ov\eta_\mu$ the set of
$4\times4$ real matrices\footnote{This somewhat strange definition of
  $\check{X}$ and the $\ov\eta$ matrices will arise naturally from the
  construction in appendix~\ref{app:k=4}.}
\begin{equation}
  \begin{split}
    \ov\eta_0=r(\sigma^0)=\mathbf{I}_{4\times4}
    \quad&;\quad
    \ov\eta_1 = r(\sigma^3) = \begin{pmatrix}
      0 & -1 & 0 & 0 \\
      1 & 0 & 0 & 0 \\
      0 & 0 & 0 & 1 \\
      0 & 0 & -1 & 0
    \end{pmatrix}
    \quad;\\
    \ov\eta_2 = -r(\sigma^2) =\begin{pmatrix}
      0 & 0 & -1 & 0 \\
      0 & 0 & 0 & -1 \\
      1 & 0 & 0 & 0 \\
      0 & 1 & 0 & 0
    \end{pmatrix}
    \quad&;\quad
    \ov\eta_3 = r(\sigma^1) = \begin{pmatrix}
      0 & 0 & 0 & -1 \\
      0 & 0 & 1 & 0 \\
      0 & -1 & 0 & 0 \\
      1 & 0 & 0 & 0
    \end{pmatrix}.
  \end{split}
\end{equation}
As stated above, this tachyon has Pontryagin class equal to two (when
defined over the sphere); it thus describes a two-center
Taub-NUT. This is the tachyon profile used in the main text to
construct the $A_1$ singular CFT. 

\medskip

For $k=4$, the Dirac matrices appearing in the  expression of the ABS tachyon are 16-dimensional,  which translates into $8$-dimensional chiral spinor bundles. The  homomorphism \eq{fgetcomp} turn these to real $16\times 16$ matrices. In 8 dimensions, one can take a Majorana-Weyl condition of spinors. If the original matrices used in the ABS tachyon are in this representation, then the homomorphism \eq{fgetcomp} will yield two irreducible 8-dimensional blocks, each of which will constitute a Majorana-Weyl representation of the $SO(8)$ Clifford algebra. This means that the soliton constructed via turning the ABS tachyon into an $SO(8)$ tachyon via \eq{fgetcomp} describes twice the generator of $KO(S^8)$, in accordance with general considerations above. In this particular case, one could directly take the ABS tachyon with real $8\times 8$ Gamma matrices and that would describe the generator of $KO(S^8)$ \cite{Witten:1998cd}:
\begin{align} T=X^\mu(\Gamma_{8\times8})_\mu.\end{align}

\medskip

Dirac matrices in $k=2$ dimensions can be taken as $\tau^1,\tau^2$. The ABS tachyon is then $\tau^1X^1+\tau^2 X^2$ which, when restricted to chiral spinors, yields simply $T_{ABS}=X^1+iX^2\equiv Z$. The real tachyon is then
\begin{align}T=\left(\begin{array}{cc}X_1&-X_2\\X_2&X_1\end{array}\right)=\vert X\vert\left(\begin{array}{cc}\cos\theta&-\sin\theta\\\sin\theta&\cos\theta\end{array}\right)\label{tachk2} .\end{align}
This tachyon has winding one around infinity. This
  winding cannot be undone with $SO(2)$ gauge transformations, but if
  embedded in $SO(n)$ for $n>2$ (in the K-theoretic spirit) the
  winding can be undone in pairs. To be more specific, the tachyon specifies a closed curve in $SO(2)$, which can be embedded in $SO(n)$. A closed curve in $SO(n)$ is contractible if and only if it lifts to a closed curve in the universal cover $Spin(n)$. A curve of winding $1$ lifts to a curve connecting the identity in $Spin(n)$ to minus the identity, whereas a curve of even winding lifts to a closed curve. Thus, $KO(S^2)=\IZ_2$ with the generator being precisely \eq{tachk2}, as desired.

\medskip

For $k=1$, the ABS construction strictly does not apply, since we are in odd dimension. Nevertheless, one can formally write an ABS tachyon $\Gamma^\mu X_\mu$, with $\Gamma^\mu$ furnishing a representation of the $SO(k)$ Clifford algebra. For $k=1$, the complex tachyon is just $T=X$. The realification prescription tells us that the real tachyon is $x\mathbf{1}_{2\times2}$. The determinant of this tachyon is $+1$ everywhere. The two classes of $KO(S^1)=\IZ_2$ correspond to orientable vector bundles and non-orientable vector bundles, which is directly measured by the sign of the determinant of the tachyon. Therefore, the tachyon constructed forcing the ABS prescription in this case does not generate $KO(S^1)$. 

However, the above discussion also suggests a solution, similar to what happened in the $k=8$ case above: simply take $T=X$ as the real tachyon. The sign of the determinant indeed changes, so this corresponds to a non-orientable bundle over $S^1$. In a sense the ABS prescription still works, only that like in the $k=8$ case it generates an order two element of the $KO$ group, which happens to vanish identically in $KO(S^1)$.

\subsection{An alternative embedding for the $k=4$ instanton}\label{app:k=4}
In the particular case of $k=4$, there is another natural route to arrive at the tachyon \eq{REALabs} in a different way, using the fact that $SO(4)$ can be double covered by $SU(2)\times SU(2)$. Along the route we will develop technology which will be essential for the arguments in section~\ref{sec:cft}. 

Using the $\sigma$ matrices~\eqref{eq:sigma-matrices} we can construct
a bijection between 4-vectors $V^\mu$ of $SO(4)$ and bispinors
$\sV_{\alpha\dot{\alpha}}$ (we will omit the subindices for
conciseness, and call the bivector simply $\sV$), given by
\begin{align}
  \sV & = \sigma^\mu V^\mu\\
  V^\mu & = \frac{1}{2}\tr(\ov\sigma^\mu\sV)\, .
\end{align}
We take a summation convention in which any repeated index,
independent of whether it is ``up'' or ``down'', is to be summed
over. A generic element $(g_1,g_2)=(e^{il_k\tau^k}, e^{ir_k \tau^k})$
of $SU(2)\times SU(2)$ acts on $\sV$ as
\begin{equation}
  \sV \to g_1 \sV g_2^{-1} = \sV + i (l_k \tau^k \sV - r_k \sV \tau^k)
  + \ldots\, .
\end{equation}
Acting on the vector representation this is
\begin{equation}
  \begin{split}
    V^\mu & \to \frac{1}{2}\tr \left(\ov\sigma^\mu g_1 \sV
      g_2^{-1}\right)\\
    & = V^\mu + \frac{i}{2} \left[l_i tr\bigl(\ov\sigma^\mu \tau^i
      \sigma^\nu \bigr) - r_i \tr\bigl(\sigma^\nu \tau^i \ov\sigma^\mu
      \bigr) + \ldots\right]V^\nu\, .
  \end{split}
\end{equation}
It is now convenient to introduce the 't Hooft matrices
\begin{align}
  \ov\eta_i^{\mu\nu} & = \frac{i}{2}\tr \bigl(\ov\sigma^\mu \tau^i
      \sigma^\nu \bigr)\, ,\\
  \eta_i^{\mu\nu} & = \frac{i}{2}\tr\bigl(\sigma^\nu \tau^i
                    \ov\sigma^\mu\bigr)\, .
\end{align}
From the definition it is clear that these are real matrices. Explicit
expressions are
\begin{align}
  \ov\eta_1 = \begin{pmatrix}
    0 & -1 & 0 & 0 \\
    1 & 0 & 0 & 0 \\
    0 & 0 & 0 & 1 \\
    0 & 0 & -1 & 0
  \end{pmatrix}
  \quad;\quad
  \ov\eta_2 = \begin{pmatrix}
    0 & 0 & -1 & 0 \\
    0 & 0 & 0 & -1 \\
    1 & 0 & 0 & 0 \\
    0 & 1 & 0 & 0
  \end{pmatrix}
  \quad;\quad
  \ov\eta_3 = \begin{pmatrix}
    0 & 0 & 0 & -1 \\
    0 & 0 & 1 & 0 \\
    0 & -1 & 0 & 0 \\
    1 & 0 & 0 & 0
  \end{pmatrix}\\
  \eta_1 = \begin{pmatrix}
    0 & 1 & 0 & 0 \\
    -1 & 0 & 0 & 0 \\
    0 & 0 & 0 & 1 \\
    0 & 0 & -1 & 0
  \end{pmatrix}
  \quad;\quad
  \eta_2 = \begin{pmatrix}
    0 & 0 & 1 & 0 \\
    0 & 0 & 0 & -1 \\
    -1 & 0 & 0 & 0 \\
    0 & 1 & 0 & 0
  \end{pmatrix}
  \quad;\quad
  \eta_3 = \begin{pmatrix}
    0 & 0 & 0 & 1 \\
    0 & 0 & 1 & 0 \\
    0 & -1 & 0 & 0 \\
    -1 & 0 & 0 & 0
  \end{pmatrix} 
\end{align}
from which we see that these matrices are antisymmetric, and one can
also check that $\ov\eta_i$ is anti-self-dual, and $\eta_i$ is
self-dual. In this way, the mapping of $SU(2)\times SU(2)$ gives a natural construction for the $\eta$ matrices showing up in \eq{REALabs}, while also providing expressions in terms of $\sigma$ matrices which are essential in section~\ref{sec:cft}. In terms of these matrices we have
\begin{equation}
  V^\mu \to V^\mu + \left(l_i \ov\eta_i^{\mu\nu} + r_i \eta_i^{\mu\nu} + \ldots\right)V^\nu\, .
\end{equation}
This provides the desired embedding of the $SU(2)\times SU(2)$ element
$(l_i\tau^i, r_i\tau^i)$ into the Lie algebra of $SO(4)$, which is
generated by the 6 real antisymmetric matrices $\ov\eta$, $\eta$. More
explicitly, a finite $SO(4)$ rotation $\Lambda$ can be written as
\begin{equation}
  \Lambda = e^{l_i\ov\eta_i + r_i\eta_i} =
  e^{l_i\ov\eta_i}e^{r_i\eta_i} = e^{r_i\eta_i}e^{l_i\ov\eta_i}
\end{equation}
using that $[\ov\eta_i,\eta_j]=0$. A last couple of relations that
will be useful are
\begin{equation}
  [\ov\eta_i, \ov\eta_j] = -2\varepsilon_{ijk}\ov\eta_k\qquad;\qquad
  [\eta_i, \eta_j] = -2\varepsilon_{ijk}\eta_k\, .
\end{equation}

The ABS $SU(2)$ tachyon determines a representative of the generator of $K(S^4)$. The gauge connection in a representative of the form $(E,0)$ of this K-theory class must be the famous BPST $SU(2)$ instanton. In regular gauge, this can be written as
\begin{equation}
  \label{eq:SU(2)-instanton}
  A_\mu = \frac{x_\nu}{x^2 + \rho^2}  \ov\eta_i^{\mu\nu} \tau^i
\end{equation}
where we have fixed the instanton at the origin, denoted the instanton
size by $\rho$ and picked a particular orientation in $SU(2)$. We now
need to pick an embedding of this $SU(2)$ into $SO(4)$. Choosing to
embed it in the first factor, we identify
$l_i=\ov\eta^{\mu\nu}_i$. From the discussion above, the self-dual
$SO(4)$ instanton in regular gauge is then given by
\begin{equation}
  \label{eq:SO(4)-instanton}
  A_\mu = \frac{x_\nu}{x^2+\rho^2} G^{\mu\nu}
\end{equation}
with $G^{\mu\nu}$ a set of 6 $SO(4)$ matrices of the form
\begin{align}
  (G^{\mu\nu})^{\alpha\beta} = \ov\eta_i^{\mu\nu}
  \ov\eta_i^{\alpha\beta}\, .
\end{align}

Finite-energy considerations require that the connection must be pure
gauge at infinity
\begin{equation}
  A_\mu \to U^{-1}\partial_\mu U + \ldots
\end{equation}
with $U$ a gauge transformation. In the case of the $SU(2)$ instanton
we have that \cite{Vandoren:2008xg}
\begin{equation}
  \label{eq:SU(2)-asymptotic-connection}
  U^{-1}\partial_\mu U = -\sigma^{\mu\nu} \frac{x_\nu}{x^2}
\end{equation}
with $\sigma^{\mu\nu}=i\ov\eta_i^{\mu\nu}\tau^i$ the generator of the
Lorentz group in the spinorial representation. The generalization to
the vector embedding is straightforward, we have
\begin{equation}
  U^{-1}\partial_\mu U = G^{\mu\nu} \frac{x_\nu}{x^2}\, .
\end{equation}
In the $SU(2)$ case we can solve for $U$
in~\eqref{eq:SU(2)-asymptotic-connection}, obtaining 
\begin{align}
  U = \frac{\sigma^\mu x_\mu}{\sqrt{x^2}}\equiv \hat{x}_\mu
  \sigma^\mu\, .
\end{align}
What makes this rewriting possible is the following familiar fact. Any
$SU(2)=S^3$ element can be specified by three numbers $\lambda_i$,
with $U=\exp(i\lambda_i\tau^i)$. But it can also be expressed as a
unit quaternion $U=\beta_\mu\sigma^\mu$, as long as
$\beta\cdot\beta=1$. This gives a natural one-to-one map between the
3-sphere at infinity and $SU(2)$ (in the spinor representation).

The generalization of the exponential representation to the vector
representation is straightforward
\begin{equation}
  U = \exp(i\lambda_i \ov\eta_i)
\end{equation}
with the $\lambda_i$ determined from $x_\mu$ as before. For the
generalization of the quaternion representation, rewrite
\begin{equation}
  \mathsf{x} = \hat{x}_\mu \sigma^\mu\, .
\end{equation}
This defines an element of a $SU(2)\subset SO(4)$, as before. Acting
on any vector $\sV = V_\mu\sigma^\mu$ this acts as (in our conventions)
\begin{equation}
  \sV \to \mathsf{x}\sV
\end{equation}
which extracting components gives
\begin{equation}
  \begin{split}
    (V')^\mu & = \frac{1}{2}\tr(\ov\sigma^\mu \mathsf{x}\sV)\\
    & = \frac{1}{2}V^\rho \hat{x}^\tau \tr(\ov\sigma^\mu \sigma^\tau
    \sigma^\rho)
  \end{split}
\end{equation}
We read off the explicit mapping in this way:
\begin{equation}
  U^{\mu\rho} = \frac{1}{2}\hat{x}^\tau \tr( \ov\sigma^\mu \sigma^\tau
    \sigma^\rho)\, .
\end{equation}
Defining $\eta_0\equiv 1$, this can be neatly rewritten as
\begin{equation}
  \label{eq:vector-U}
  U = \hat{x}^\tau \ov\eta_\tau
\end{equation}
which using $\{\ov\eta_i, \ov\eta_j\}=-2\delta_{ij}$, $\eta_0^2=1$ can
be easily seen to lay in $O(4)$ (and explicit computation, or using
that the $S^3$ is generated continuously from $\hat{x}=(1,0,0,0)$,
shows that it is in $SO(4)$, i.e. $\det U = +1$).

If we want to construct the instanton based on tachyon condensation
with a non-trivial tachyon profile, finite-energy considerations
impose that $T$ is basically just $U$, given by~\eqref{eq:vector-U},
with the replacement $\hat{x}\to x$ (accounting for the fact that the
actual tachyon vacuum for brane condensation is at $T=\infty$):
\begin{equation}
  \label{eq:vector-T}
  T = x^\tau \ov\eta_\tau\, .
\end{equation}

For the supercritical closed string case we choose $X^\mu$ as our base
directions, and we view the $SO(4)\times SO(4)$ symmetry as the
rotation group acting on the $x,y$ dimensions. By analogy with the
gauge instanton case we then take a tachyon profile of the form
\begin{equation}
  \label{eq:supercritical-tachyon}
  \tach = X^\tau (\ov\eta_\tau)_{ab} x^a y^b = \frac{1}{2}\tr(\widetilde{\sx} \sX \sy)
\end{equation}
where $\sX=X_\mu\sigma^\mu$, $\sy=y_\mu\sigma^\mu$ as before, and in
order to display the symmetries of the system more manifestly, we have
introduced $\widetilde{\sx} = \sx^\dagger = \widetilde{x}_\mu
\sigma^\mu$, with $\widetilde{x}_\mu=(x_0, -x_1, -x_2, -x_3)$.

\paragraph{Restriction to a spinor representation.} We have focused in
the case of $SO(4)$, so $\sV$ is naturally a bispinor. If we are
interested just in one $SU(2)$ subgroup, for example the one acting on
the left, we can think of $\sV$ as a couple of spinors $\sV^1$ and
$\sV^2$, by taking $(\sV^i)_\alpha=\sV_{\alpha i}$. Under a $SU(2)$
transformation in the left subgroup $\sV\to g\sV$, which implies
$\sV^i\to g\sV^i$. Looking to the explicit form of the $\sigma^\mu$
matrices, we have $\sV^1=(x_0+ix_3, ix_1 - x_2)$ and
$\sV^2=(ix_1+x_2, x_0 - ix_3)$. One may now show that the tachyon
constructed by embedding an $SU(2)$ instanton into $SO(4)$ comes just
from taking the real part of the ABS construction. The $SO(4)$ tachyon
profile is
\begin{equation}
  \tach=X^\tau(\bar{\eta}_\tau)_{ab}x^a y^b=\frac{1}{2}
  \text{tr}\left[(x^a\bar{\sigma}^a)(X^\tau
    \sigma^\tau)(y^b\sigma^b)\right]=\frac{1}{2}
  \text{tr}\left[(x^a\bar{\sigma}^a)T_{ABS}(x^b\sigma^b)\right].
  \label{tachprod1}
\end{equation}
Now, as above, we may regard the bispinors $(x^b\sigma^b)$ as a pair of $SU(2)$ spinors so that, as a matrix,
\begin{equation}
  (y^b\sigma^b)=(\sV^1_y \quad \sV^2_y),\quad
  (x^a\bar{\sigma}^a)=(\sV^1_x \quad \sV^2_x)^\dagger
  = \begin{pmatrix}
    (\sV^1_x)^\dagger\\
    (\sV^2_x)^\dagger
    \end{pmatrix},
\end{equation}
where $\sV^i_x,\sV^i_y$ are the column vectors defined above. Now
expanding the product in \eq{tachprod1}, we get
\begin{equation}
(x^a\bar{\sigma}^a)T_{ABS}(y^b\sigma^b)=
\begin{pmatrix}
  (\sV_x^1)^\dagger T_{ABS}\sV^1_y
  & (\sV_x^1)^\dagger T_{ABS} \sV^2_y\\
  (\sV_x^2)^\dagger T_{ABS} \sV^1_y &
  (\sV_x^2)^\dagger T_{ABS} \sV^2_y\end{pmatrix}\, .
\end{equation}
Now, since (assuming the $X^\mu$ are real as well)
\begin{equation}
  \begin{split}
    \bigl[(\sV_x^1)^\dagger T_{ABS} \sV^1_y\bigr]^* & = (\sV_x^1)^T T_{ABS}^*
    (\sV^1_y)^* = ((\sV^1_x)^T \tau^2)T_{ABS}(\tau^2 (\sV^1_y)^*) \\
    & = (\sV_x^2)^\dagger T_{ABS} \sV_y^2
  \end{split}
\end{equation}
using that
\begin{equation}
  \label{eq:sigma-conjugation}
  \sigma_\mu^* = \tau^2 \sigma_\mu \tau^2 \qquad ;\qquad
  \ov\sigma_\mu^* = \tau^2\ov\sigma_\mu\tau^2\, .
\end{equation}
and $\tau^2(\sV^1)^*=i\sV^2$. This means that~\eqref{tachprod1} can
be written as
\begin{equation}
  \tach= \re\bigl((\sV_x^1)^\dagger T_{ABS} \sV^1_y\bigr).
\end{equation}
where $\sV_x^1$ and $\sV_y^1$ are unconstrained Weyl spinors. In other
words, the real tachyon is just the real part of the complex ABS
tachyon.

\section{The caloron solution}
\label{app:caloron}

The ABS construction discussed in the main text, together with its
real version, provides us with a tachyon describing the $\IC^2/\IZ_2$
orbifold. Via deformation we obtain a two-center Taub-NUT
solution. The standard multi-center Taub-NUT metric
\cite{Ortin:2004ms} has an asymptotic circle at infinity, whose radius
is formally infinite in the $\IC^2/\IZ_2$ orbifold and its
deformations. Nevertheless, the theory at finite radius has very
interesting properties which disappear in the infinite radius
limit. For instance, the system is T-dual to a configuration of two
coincident NS5-branes.

It is interesting to think about what happens to the ABS tachyon when we take one of the
dimensions periodic. We still expect a $SO(4)$ instanton to exist in
this case, but clearly~\eqref{eq:SU(2)-instanton} will not do, since
it is not periodic. This problem was solved
by~\cite{Harrington:1978ve}, who dubbed the solution they found
``caloron''. The basic idea is simple: one uses the solution for
instantons along a line (see for instance \cite{Witten:1976ck}), taken
in the limit where there are infinite instantons along the direction
to be compactified, with periodicity equal to the compactification
radius. The resulting solution has the right periodicity.

In our case, we are directly interested in the tachyon profile, so it will be convenient to write $A=(\partial T) T^{-1}$. The caloron recipe instructs us to sum over an infinite periodic array of solutions of the form
\begin{align}A_n=(\partial T_n) T_n^{-1},\quad T_n=\frac{(\tau-\tau_0+n)\mathbf{I}+i\vec{\tau}\cdot\vec{X}}{\sqrt{(\tau-\tau_0+n)^2+\vert\vec{X}\vert^2}},\label{calor20}\end{align}
where $\tau$ is the soon-to-be compact direction and $\vec{\tau}$ the usual vector of Pauli matrices. $\tau_0$ labels the position of the caloron in the compact circle. We are interested in a periodic, pure gauge configuration (except at the locus where the tachyon is not invertible), such that the integral of the Chern-Simons three form around each point of the form $(\tau_0+n,\vec{0})$ is one. Precisely these properties are satisfied by the configuration
 \begin{align}A^c=\sum_{n=-\infty}^\infty \left(\prod_{k<n} T_k\right)A_n  \left(\prod_{k<n} T_k\right)^{-1}=(\partial T^c) (T^c)^{-1},\quad T^c=\prod_{n=-\infty}^\infty T_n.\end{align}
Thus, the tachyon we are looking for is precisely  $T^c$. It is easy to evaluate it by noting from \eq{calor20} that, given $\vec{X}$, the matrices $\mathbf{I}, i\vec{\tau}\cdot\vec{X}/\vert\vec{X}\vert$ form a representation of the algebra of the complex numbers. This turns the definition of $T^c$ into an infinite product of complex numbers of norm 1, whose argument $\theta^c$ is readily evaluated as follows. Since we are only interested in the argument, we can compute it changing the normalization of the $T_n$. Since 
 \begin{equation}T_n= \frac{n+z}{\vert n+z\vert},\quad  \text{with}\quad z\equiv \tau-\tau_0+i\vert\vec{X}\vert,\end{equation}
 we may write
 \begin{equation}\theta^c=\arg\left[\prod_{n=-\infty}^\infty \frac{n+z}{\vert n+z\vert}\right]=\arg\left[z\prod_{n=1}^\infty\left( \frac{n+z}{n}\right)\left( \frac{-n+z}{n}\right)\right]=\arg\left[z\prod_{n=1}^\infty \left(1-\frac{z^2}{n^2}\right)\right],\label{calor2}\end{equation}
 where in the last line we have changed the sign of every term in the
 product. This may introduce a sign in the final result for $T_c$
 which is in any case irrelevant, as we will argue below. The product
 in the last term is the Weierstrass product form of
 $\sin(\pi z)/(\pi z)$, so that
 \begin{equation}\theta^c=\arg\left[\frac{\sin(\pi z)}{\pi}\right].\label{thcfinal}\end{equation}
 From this, one can construct the periodic caloron tachyon
\begin{equation}
  \begin{split}
    T^c&=e^{i\theta^c\vec{\tau}\cdot\frac{\vec{X}}{\vert\vec{X}\vert}}=\cos(\theta^c)\,\mathbf{I}+i\sin(\theta^c)\,\vec{\tau}\cdot\frac{\vec{X}}{\vert\vec{X}\vert}\\
    & =
    \frac{\tan(\pi(\tau-\tau_0))\,\mathbf{I}+i\tanh\bigl(\pi|\vec{X}|\bigr)\,\vec{\tau}\cdot\frac{\vec{X}}{\vert\vec{X}\vert}}{\left[\tanh^2\bigl(\pi
        |\vec{X}|\bigr) +
        \tan^2(\pi(\tau-\tau_0))\right]^{\frac{1}{2}}} = \frac{T^c_\mu}{|T^c_\mu|}\sigma^\mu.\label{calor3}
  \end{split}
\end{equation}
with
$T^c_\mu=\left\{\tan(\pi(\tau-\tau_0)),\tanh\bigl(\pi|\vec{X}|\bigr)\frac{\vec{X}}{\vert\vec{X}\vert}\right\}$. In
writing the expression in terms of tangents one needs to make a choice
of branch cut in the square root in~\eqref{calor3}, we have chosen the
sign that agrees with~\eqref{thcfinal} on
$|\tau-\tau_0|\leq \frac{1}{2}$. Relatedly, notice that
$\tau\to\tau+1$ sends $\theta_c\to \theta_c+\pi$, or equivalently
$T^c\to -T^c$. When we view the caloron as a configuration in $SO(4)$
Yang-Mills theory this is fine, since the physical object, the
connection $A^c$, is invariant under this transformation. In the
context of the supercritical string the tachyon $T^c$ is physical,
transforming in a bifundamental of $SO(4)\times SO(4)$, and
$T^c\to -T^c$ is simply a gauge transformation in the center of the
gauge group.


An interesting point is that on general grounds we expect the endpoint of the condensation with tachyon \eq{calor3} to be an ordinary two-center Taub-NUT space. This solution has a $U(1)$ isometry along the compact direction which appears to be broken explicitly \eq{calor3}. Indeed, \eq{calor2} and therefore \eq{calor3} have a free parameter $\tau_0$, which corresponds to moving the center of the instanton in the compact direction. Thus, changing $\tau_0$ either corresponds to an irrelevant perturbation of the CFT, or describes some deformation of the ordinary Taub-NUT space. Evidence for the latter comes from noticing that if we separate the two centers of the Taub-NUT space, the theory is no longer singular. In the UV, we could imagine separating the coincident centers of the Pontryagin number 2 solution constructed above into two configurations with Pontryagin number one. Presumably this would correspond to tuning the difference $\tau^1_0-\tau_0^2$ of the $\IS^1$ position of the two Taub-NUT centers away from zero. In the IR, the modulus which desingularizes the CFT is $B$ field on the vanishing 2-cycle of the configuration. The parameter $\tau_0=\tau^1_0+\tau_0^2$ of our tachyon profile therefore seems to correspond to turning a $B$-field on the center-of-mass normalizable two-form in the IR.

\bibliographystyle{JHEP}
\bibliography{refs}

\providecommand{\href}[2]{#2}\begingroup\raggedright\begin{thebibliography}{10}

\bibitem{Sen:1998rg}
A.~Sen, {\it {Stable nonBPS states in string theory}},  {\em JHEP} {\bf 9806}
  (1998) 007, [\href{http://xxx.lanl.gov/abs/hep-th/9803194}{{\tt
  hep-th/9803194}}].

\bibitem{Sen:1999mg}
A.~Sen, {\it {NonBPS states and Branes in string theory}},
  \href{http://xxx.lanl.gov/abs/hep-th/9904207}{{\tt hep-th/9904207}}.

\bibitem{Witten:1998cd}
E.~Witten, {\it {D-branes and K theory}},  {\em JHEP} {\bf 9812} (1998) 019,
  [\href{http://xxx.lanl.gov/abs/hep-th/9810188}{{\tt hep-th/9810188}}].

\bibitem{Moore:1999gb}
G.~W. Moore and E.~Witten, {\it {Selfduality, Ramond-Ramond fields, and K
  theory}},  {\em JHEP} {\bf 0005} (2000) 032,
  [\href{http://xxx.lanl.gov/abs/hep-th/9912279}{{\tt hep-th/9912279}}].

\bibitem{Collinucci:2008pf}
A.~Collinucci, F.~Denef, and M.~Esole, {\it {D-brane Deconstructions in IIB
  Orientifolds}},  {\em JHEP} {\bf 0902} (2009) 005,
  [\href{http://xxx.lanl.gov/abs/0805.1573}{{\tt arXiv:0805.1573}}].

\bibitem{Collinucci:2014qfa}
A.~Collinucci and R.~Savelli, {\it {T-branes as branes within branes}},
  \href{http://xxx.lanl.gov/abs/1410.4178}{{\tt arXiv:1410.4178}}.

\bibitem{Hellerman:2004zm}
S.~Hellerman, {\it {On the landscape of superstring theory in $D > 10$}},
  \href{http://xxx.lanl.gov/abs/hep-th/0405041}{{\tt hep-th/0405041}}.

\bibitem{Hellerman:2004qa}
S.~Hellerman and X.~Liu, {\it {Dynamical dimension change in supercritical
  string theory}},  \href{http://xxx.lanl.gov/abs/hep-th/0409071}{{\tt
  hep-th/0409071}}.

\bibitem{Hellerman:2006ff}
S.~Hellerman and I.~Swanson, {\it {Dimension-changing exact solutions of string
  theory}},  {\em JHEP} {\bf 0709} (2007) 096,
  [\href{http://xxx.lanl.gov/abs/hep-th/0612051}{{\tt hep-th/0612051}}].

\bibitem{Garcia-Etxebarria:2014txa}
I.~García-Etxebarria, M.~Montero, and A.~Uranga, {\it {Heterotic NS5-branes
  from closed string tachyon condensation}},  {\em Phys.Rev.} {\bf D90} (2014),
  no.~12 126002, [\href{http://xxx.lanl.gov/abs/1405.0009}{{\tt
  arXiv:1405.0009}}].

\bibitem{Berasaluce-Gonzalez:2013sna}
M.~Berasaluce-González, M.~Montero, A.~Retolaza, and A.~M. Uranga, {\it
  {Discrete gauge symmetries from (closed string) tachyon condensation}},  {\em
  JHEP} {\bf 1311} (2013) 144, [\href{http://xxx.lanl.gov/abs/1305.6788}{{\tt
  arXiv:1305.6788}}].

\bibitem{Braun:2011zm}
A.~P. Braun, A.~Collinucci, and R.~Valandro, {\it {G-flux in F-theory and
  algebraic cycles}},  {\em Nucl.Phys.} {\bf B856} (2012) 129--179,
  [\href{http://xxx.lanl.gov/abs/1107.5337}{{\tt arXiv:1107.5337}}].

\bibitem{Braun:2012nk}
A.~P. Braun, A.~Collinucci, and R.~Valandro, {\it {Algebraic description of
  G-flux in F-theory: new techniques for F-theory phenomenology}},  {\em
  Fortsch.Phys.} {\bf 60} (2012) 934--940,
  [\href{http://xxx.lanl.gov/abs/1202.5029}{{\tt arXiv:1202.5029}}].

\bibitem{Collinucci:2014taa}
A.~Collinucci and R.~Savelli, {\it {F-theory on singular spaces}},
  \href{http://xxx.lanl.gov/abs/1410.4867}{{\tt arXiv:1410.4867}}.

\bibitem{Chamseddine:1991qu}
A.~H. Chamseddine, {\it {A Study of noncritical strings in arbitrary
  dimensions}},  {\em Nucl.Phys.} {\bf B368} (1992) 98--120.

\bibitem{Dixon:1986iz}
L.~J. Dixon and J.~A. Harvey, {\it {String Theories in Ten-Dimensions Without
  Space-Time Supersymmetry}},  {\em Nucl.Phys.} {\bf B274} (1986) 93--105.

\bibitem{Seiberg:1986by}
N.~Seiberg and E.~Witten, {\it {Spin Structures in String Theory}},  {\em
  Nucl.Phys.} {\bf B276} (1986) 272.

\bibitem{Bergman:1999km}
O.~Bergman and M.~R. Gaberdiel, {\it {Dualities of type 0 strings}},  {\em
  JHEP} {\bf 9907} (1999) 022,
  [\href{http://xxx.lanl.gov/abs/hep-th/9906055}{{\tt hep-th/9906055}}].

\bibitem{Hellerman:2006nx}
S.~Hellerman and I.~Swanson, {\it {Cosmological solutions of supercritical
  string theory}},  {\em Phys.Rev.} {\bf D77} (2008) 126011,
  [\href{http://xxx.lanl.gov/abs/hep-th/0611317}{{\tt hep-th/0611317}}].

\bibitem{Hellerman:2006hf}
S.~Hellerman and I.~Swanson, {\it {Cosmological unification of string
  theories}},  {\em JHEP} {\bf 0807} (2008) 022,
  [\href{http://xxx.lanl.gov/abs/hep-th/0612116}{{\tt hep-th/0612116}}].

\bibitem{Hellerman:2007fc}
S.~Hellerman and I.~Swanson, {\it {Charting the landscape of supercritical
  string theory}},  {\em Phys.Rev.Lett.} {\bf 99} (2007) 171601,
  [\href{http://xxx.lanl.gov/abs/0705.0980}{{\tt arXiv:0705.0980}}].

\bibitem{Hellerman:2007zz}
S.~Hellerman and I.~Swanson, {\it {A Stable vacuum of the tachyonic E(8)
  string}},  \href{http://xxx.lanl.gov/abs/0710.1628}{{\tt arXiv:0710.1628}}.

\bibitem{Hellerman:2007ym}
S.~Hellerman and I.~Swanson, {\it {Supercritical N = 2 string theory}},
  \href{http://xxx.lanl.gov/abs/0709.2166}{{\tt arXiv:0709.2166}}.

\bibitem{Hellerman:2008wp}
S.~Hellerman and M.~Schnabl, {\it {Light-like tachyon condensation in Open
  String Field Theory}},  {\em JHEP} {\bf 1304} (2013) 005,
  [\href{http://xxx.lanl.gov/abs/0803.1184}{{\tt arXiv:0803.1184}}].

\bibitem{Sen:2004nf}
A.~Sen, {\it {Tachyon dynamics in open string theory}},  {\em Int.J.Mod.Phys.}
  {\bf A20} (2005) 5513--5656,
  [\href{http://xxx.lanl.gov/abs/hep-th/0410103}{{\tt hep-th/0410103}}].

\bibitem{Atiyah:1964zz}
M.~Atiyah, R.~Bott, and A.~Shapiro, {\it {Clifford modules}},  {\em Topology}
  {\bf 3} (1964) S3--S38.

\bibitem{Bergman:1997rf}
O.~Bergman and M.~R. Gaberdiel, {\it {A Nonsupersymmetric open string theory
  and S duality}},  {\em Nucl.Phys.} {\bf B499} (1997) 183--204,
  [\href{http://xxx.lanl.gov/abs/hep-th/9701137}{{\tt hep-th/9701137}}].

\bibitem{Green:1996dd}
M.~B. Green, J.~A. Harvey, and G.~W. Moore, {\it {I-brane inflow and anomalous
  couplings on d-branes}},  {\em Class.Quant.Grav.} {\bf 14} (1997) 47--52,
  [\href{http://xxx.lanl.gov/abs/hep-th/9605033}{{\tt hep-th/9605033}}].

\bibitem{Cheung:1997az}
Y.-K.~E. Cheung and Z.~Yin, {\it {Anomalies, branes, and currents}},  {\em
  Nucl.Phys.} {\bf B517} (1998) 69--91,
  [\href{http://xxx.lanl.gov/abs/hep-th/9710206}{{\tt hep-th/9710206}}].

\bibitem{Minasian:1997mm}
R.~Minasian and G.~W. Moore, {\it {K theory and Ramond-Ramond charge}},  {\em
  JHEP} {\bf 9711} (1997) 002,
  [\href{http://xxx.lanl.gov/abs/hep-th/9710230}{{\tt hep-th/9710230}}].

\bibitem{Witten:1994tz}
E.~Witten, {\it {Sigma models and the ADHM construction of instantons}},  {\em
  J.Geom.Phys.} {\bf 15} (1995) 215--226,
  [\href{http://xxx.lanl.gov/abs/hep-th/9410052}{{\tt hep-th/9410052}}].

\bibitem{Witten:1995zh}
E.~Witten, {\it {Some comments on string dynamics}},
  \href{http://xxx.lanl.gov/abs/hep-th/9507121}{{\tt hep-th/9507121}}.

\bibitem{Vafa:1995fj}
C.~Vafa and E.~Witten, {\it {A One loop test of string duality}},  {\em
  Nucl.Phys.} {\bf B447} (1995) 261--270,
  [\href{http://xxx.lanl.gov/abs/hep-th/9505053}{{\tt hep-th/9505053}}].

\bibitem{Witten:1995em}
E.~Witten, {\it {Five-branes and M theory on an orbifold}},  {\em Nucl.Phys.}
  {\bf B463} (1996) 383--397,
  [\href{http://xxx.lanl.gov/abs/hep-th/9512219}{{\tt hep-th/9512219}}].

\bibitem{Forte:1986em}
S.~Forte, {\it {Explicit Construction of Anomalies}},  {\em Nucl.Phys.} {\bf
  B288} (1987) 252.

\bibitem{nakahara2003geometry}
M.~Nakahara, {\em Geometry, Topology and Physics, Second Edition}.
\newblock Graduate student series in physics. Taylor \& Francis, 2003.

\bibitem{Liu:2013dna}
J.~T. Liu and R.~Minasian, {\it {Higher-derivative couplings in string theory:
  dualities and the $B$-field}},  {\em Nucl.Phys.} {\bf B874} (2013) 413--470,
  [\href{http://xxx.lanl.gov/abs/1304.3137}{{\tt arXiv:1304.3137}}].

\bibitem{Gukov:1999yn}
S.~Gukov, {\it {K theory, reality, and orientifolds}},  {\em Commun.Math.Phys.}
  {\bf 210} (2000) 621--639,
  [\href{http://xxx.lanl.gov/abs/hep-th/9901042}{{\tt hep-th/9901042}}].

\bibitem{Witten:1993yc}
E.~Witten, {\it {Phases of N=2 theories in two-dimensions}},  {\em Nucl.Phys.}
  {\bf B403} (1993) 159--222,
  [\href{http://xxx.lanl.gov/abs/hep-th/9301042}{{\tt hep-th/9301042}}].

\bibitem{Doran:2007jw}
C.~Doran, B.~Greene, and S.~Judes, {\it {Families of quintic Calabi-Yau 3-folds
  with discrete symmetries}},  {\em Commun.Math.Phys.} {\bf 280} (2008)
  675--725, [\href{http://xxx.lanl.gov/abs/hep-th/0701206}{{\tt
  hep-th/0701206}}].

\bibitem{2005JGP....53...31H}
T.~{H{\"u}bsch} and A.~{Rahman}, {\it {On the geometry and homology of certain
  simple stratified varieties}},  {\em Journal of Geometry and Physics} {\bf
  53} (Jan., 2005) 31--48, [\href{http://xxx.lanl.gov/abs/math/0210394}{{\tt
  math/0210394}}].

\bibitem{Strominger:1995cz}
A.~Strominger, {\it {Massless black holes and conifolds in string theory}},
  {\em Nucl.Phys.} {\bf B451} (1995) 96--108,
  [\href{http://xxx.lanl.gov/abs/hep-th/9504090}{{\tt hep-th/9504090}}].

\bibitem{Klemm:1996hh}
A.~Klemm, P.~Mayr, and C.~Vafa, {\it {BPS states of exceptional noncritical
  strings}},  {\em Nucl.Phys.Proc.Suppl.} {\bf 58} (1997) 177,
  [\href{http://xxx.lanl.gov/abs/hep-th/9607139}{{\tt hep-th/9607139}}].

\bibitem{Aldazabal:1996du}
G.~Aldazabal, A.~Font, L.~E. Ibanez, and A.~Uranga, {\it {New branches of
  string compactifications and their F theory duals}},  {\em Nucl.Phys.} {\bf
  B492} (1997) 119--151, [\href{http://xxx.lanl.gov/abs/hep-th/9607121}{{\tt
  hep-th/9607121}}].

\bibitem{Braun:2014oya}
V.~Braun and D.~R. Morrison, {\it {F-theory on Genus-One Fibrations}},  {\em
  JHEP} {\bf 1408} (2014) 132, [\href{http://xxx.lanl.gov/abs/1401.7844}{{\tt
  arXiv:1401.7844}}].

\bibitem{Morrison:2014era}
D.~R. Morrison and W.~Taylor, {\it {Sections, multisections, and U(1) fields in
  F-theory}},  \href{http://xxx.lanl.gov/abs/1404.1527}{{\tt arXiv:1404.1527}}.

\bibitem{Anderson:2014yva}
L.~B. Anderson, I.~García-Etxebarria, T.~W. Grimm, and J.~Keitel, {\it
  {Physics of F-theory compactifications without section}},  {\em JHEP} {\bf
  1412} (2014) 156, [\href{http://xxx.lanl.gov/abs/1406.5180}{{\tt
  arXiv:1406.5180}}].

\bibitem{Klevers:2014bqa}
D.~Klevers, D.~K. Mayorga~Pena, P.-K. Oehlmann, H.~Piragua, and J.~Reuter, {\it
  {F-Theory on all Toric Hypersurface Fibrations and its Higgs Branches}},
  {\em JHEP} {\bf 1501} (2015) 142,
  [\href{http://xxx.lanl.gov/abs/1408.4808}{{\tt arXiv:1408.4808}}].

\bibitem{Garcia-Etxebarria:2014qua}
I.~García-Etxebarria, T.~W. Grimm, and J.~Keitel, {\it {Yukawas and discrete
  symmetries in F-theory compactifications without section}},  {\em JHEP} {\bf
  1411} (2014) 125, [\href{http://xxx.lanl.gov/abs/1408.6448}{{\tt
  arXiv:1408.6448}}].

\bibitem{Mayrhofer:2014haa}
C.~Mayrhofer, E.~Palti, O.~Till, and T.~Weigand, {\it {Discrete Gauge
  Symmetries by Higgsing in four-dimensional F-Theory Compactifications}},
  {\em JHEP} {\bf 1412} (2014) 068,
  [\href{http://xxx.lanl.gov/abs/1408.6831}{{\tt arXiv:1408.6831}}].

\bibitem{Mayrhofer:2014laa}
C.~Mayrhofer, E.~Palti, O.~Till, and T.~Weigand, {\it {On Discrete Symmetries
  and Torsion Homology in F-Theory}},
  \href{http://xxx.lanl.gov/abs/1410.7814}{{\tt arXiv:1410.7814}}.

\bibitem{Cvetic:2015moa}
M.~Cvetič, R.~Donagi, D.~Klevers, H.~Piragua, and M.~Poretschkin, {\it
  {F-Theory Vacua with $Z_3$ Gauge Symmetry}},
  \href{http://xxx.lanl.gov/abs/1502.0695}{{\tt arXiv:1502.0695}}.

\bibitem{Grimm:2015ona}
T.~W. Grimm, T.~G. Pugh, and D.~Regalado, {\it {Non-Abelian discrete gauge
  symmetries in F-theory}},  \href{http://xxx.lanl.gov/abs/1504.0627}{{\tt
  arXiv:1504.0627}}.

\bibitem{Donagi:2011jy}
R.~Donagi and M.~Wijnholt, {\it {Gluing Branes, I}},  {\em JHEP} {\bf 1305}
  (2013) 068, [\href{http://xxx.lanl.gov/abs/1104.2610}{{\tt
  arXiv:1104.2610}}].

\bibitem{Herbst:2008jq}
M.~Herbst, K.~Hori, and D.~Page, {\it {Phases Of N=2 Theories In 1+1 Dimensions
  With Boundary}},  \href{http://xxx.lanl.gov/abs/0803.2045}{{\tt
  arXiv:0803.2045}}.

\bibitem{Vandoren:2008xg}
S.~Vandoren and P.~van Nieuwenhuizen, {\it {Lectures on instantons}},
  \href{http://xxx.lanl.gov/abs/0802.1862}{{\tt arXiv:0802.1862}}.

\bibitem{Ortin:2004ms}
T.~Ort\'{i}n, {\em Gravity and strings}.
\newblock Cambridge University Press, 1st edition~ed., 2007.

\bibitem{Harrington:1978ve}
B.~J. Harrington and H.~K. Shepard, {\it {Periodic Euclidean Solutions and the
  Finite Temperature Yang-Mills Gas}},  {\em Phys.Rev.} {\bf D17} (1978) 2122.

\bibitem{Witten:1976ck}
E.~Witten, {\it {Some Exact Multi - Instanton Solutions of Classical Yang-Mills
  Theory}},  {\em Phys.Rev.Lett.} {\bf 38} (1977) 121--124.

\end{thebibliography}\endgroup

\end{document}